\documentclass[fleqn,usenatbib,useAMS]{./mnras}

\usepackage{newtxtext,newtxmath}
\usepackage{comment}

\usepackage[T1]{fontenc}
\usepackage{ae,aecompl}
\usepackage{comment}

\usepackage{graphicx}	
\usepackage{adjustbox}
\usepackage{color}

\newcommand{\kms}{km~s$^{-1}$}

\newcommand{\spitzer}{\textit{Spitzer}}
\newcommand{\herschel}{\textit{Herschel}}

\newcommand{\siiif}{\ion{[S}{III}]}

\newcommand{\siliif}{\ion{[Si}{II}]}

\newcommand{\mic}{$\mu$m}

\newcommand{\new}{\textcolor{black}}

\usepackage{xcolor}
\definecolor{PineGreen}{HTML}{008B72}

\title[SOFIA/HAWC+ Polarization in Cas~A]{Far-infrared Polarization of the Supernova Remnant Cassiopeia A with SOFIA HAWC+}

\author[Rho, et al.]{
Jeonghee Rho,$^1$
Aravind P. Ravi,$^2$
Le Ngoc Tram,$^3$
Thiem Hoang,$^4$
J\'er\'emy Chastenet,$^5$ \newauthor
Matthew Millard,$^{2,6}$ 
Michael J. Barlow,$^7$
Ilse De Looze,$^{5,7}$ 
Haley L. Gomez,$^8$ 
\newauthor
Florian Kirchschlager,$^{5,7}$ 
Loretta Dunne$^8$
\\
$^1$SETI Institute, 339 N. Bernardo Ave., Ste. 200, Mountain View, CA 94043; jrho@seti.org \\
$^2$Box 19059, Department of Physics, University of Texas at Arlington, Arlington, TX 76019 \\
$^3$Max Planck Institute for Radio Astronomy, Bonn, Germany\\ 
$^4$Korea Astronomy and Space Science Institute (KASI)\\
$^5$Sterrenkundig Observatorium, Ghent University, Krijgslaan 281-S9, 9000 Gent, Belgium \\
$^6$Department of Physics and Astronomy, University of Iowa, Van Allen Hall, Iowa City, IA 52242, USA; matthew-j-millard@uiowa.edu \\
$^7$Dept. of Physics \& Astronomy, University College London, Gower Street, London WC1E 6BT, UK\\
$^8$Cardiff Hub for Astrophysical Research and Technology (CHART), School of Physics \& Astronomy, Cardiff University, The Parade, Cardiff, CF24 3AA, UK 
}

\pubyear{2023}

\begin{document}

\label{firstpage}
\pagerange{\pageref{firstpage}--\pageref{lastpage}}
\maketitle

\begin{abstract}

We present polarization observations of the young supernova remnant (SNR) Cas~A using the High-resolution Airborne Wideband Camera-Plus (HAWC+) instrument onboard the Stratospheric Observatory for Infrared Astronomy (SOFIA).  The polarization map at 154 \mic\ reveals dust grains with strong polarization fractions (5 - 30\,per\,cent), supporting previous measurements made over a smaller region of the remnant at 850 \mic. The 154 \mic\ emission and the polarization signal is coincident with a region of cold dust observed in the southeastern shell and in the unshocked central ejecta. 
The highly polarized far-IR emission implies the grains are 
large ($>$0.14 \mic) and silicate-dominated. The polarization level varies across the SNR, with an inverse correlation between the polarization degree and the intensity and smaller polarization angle dispersion for brighter SNR emission. Stronger polarization is detected between the bright structures. This may result from a higher collision rate between the gas and dust producing a lower grain alignment efficiency where the gas density is higher. We use the dust emission to provide an estimate of the magnetic field strength in Cas~A using the Davis-Chandrasekhar-Fermi method.  The high polarization level is direct evidence that grains are highly elongated and strongly aligned with the magnetic field of the SNR. The dust mass from the polarized region is 0.14$\pm$0.04 M$_\odot$, a lower limit of the amount of dust present within the ejecta of Cas A. 
This result strengthens the hypothesis that core-collapse SNe are an important contributor to the dust mass in high redshift galaxies.

\end{abstract}


\section{Introduction}
The large reservoirs of dust observed in some high redshift galaxies have been hypothesized to originate from dust produced by core-collapse supernovae (ccSNe). For ccSNe to account for the dust mass budget in the early Universe, 0.1-1 M$_\odot$ of metals need to condense into solid dust grains after every massive star explosion \citep{morgan03,dwek04,dunne11,rowlands14}. Theoretical studies \citep{todini01,nozawa10,nozawa11} tend to support these high condensation efficiencies, but the quantities of dust detected from supernovae during the first 1000 days after explosion have long remained two orders of magnitude below these theoretical predictions. Recent \herschel\ and SOFIA studies were able to probe the colder dust in young supernova remnants (SNRs), detecting large quantities of dust mass in Cas~A \citep{barlow10,deLooze17}, SN1987A \citep{matsuura15,matsuura18}, the Crab Nebula \citep{Gomez12,deLooze19,nehme19}, the SNR G54.1+0.3 \citep{rho18,temim17}, and N132D \citep{rho23}. The results suggest that some ccSNe are dust factories, and thus they could be important dust sources at high redshifts.

Chemical evolution models have consistently shown that AGB stars are mostly not contributing a significant mass of dust in galaxies at high redshifts \citep{morgan03,dwek07,gall11,rowlands14}. Although AGB stars with initial main sequence masses greater than 5 M$_\odot$ could reach their dust production phase (Asymptotic Giant Branch) in 50 Myr, it is unclear how much dust is formed in these stars 
\citep{morgan03}. An alternative route to dust formation in the early Universe is via grain growth in the interstellar medium (ISM) \citep{zhukovska18}, seeded by heavy elements and/or surviving dust grains from SNe \citep{draine09p}. However, there are no convincing studies to support dust growth in the ISM that can explain dust in high-redshift galaxies. 

The appearance of large quantities of dust at high redshifts could be explained if ccSNe do not subsequently destroy most of the dust they create \citep{michalowski15}. Unfortunately poor constraints on the grain size and composition of newly condensed dust in SNRs results in large uncertainties in the mass of dust formed and the fraction that could survive the reverse shock \citep{kirchschlager19}.  

\begin{figure*}
\includegraphics[scale=0.6,angle=0,width=17.3truecm]{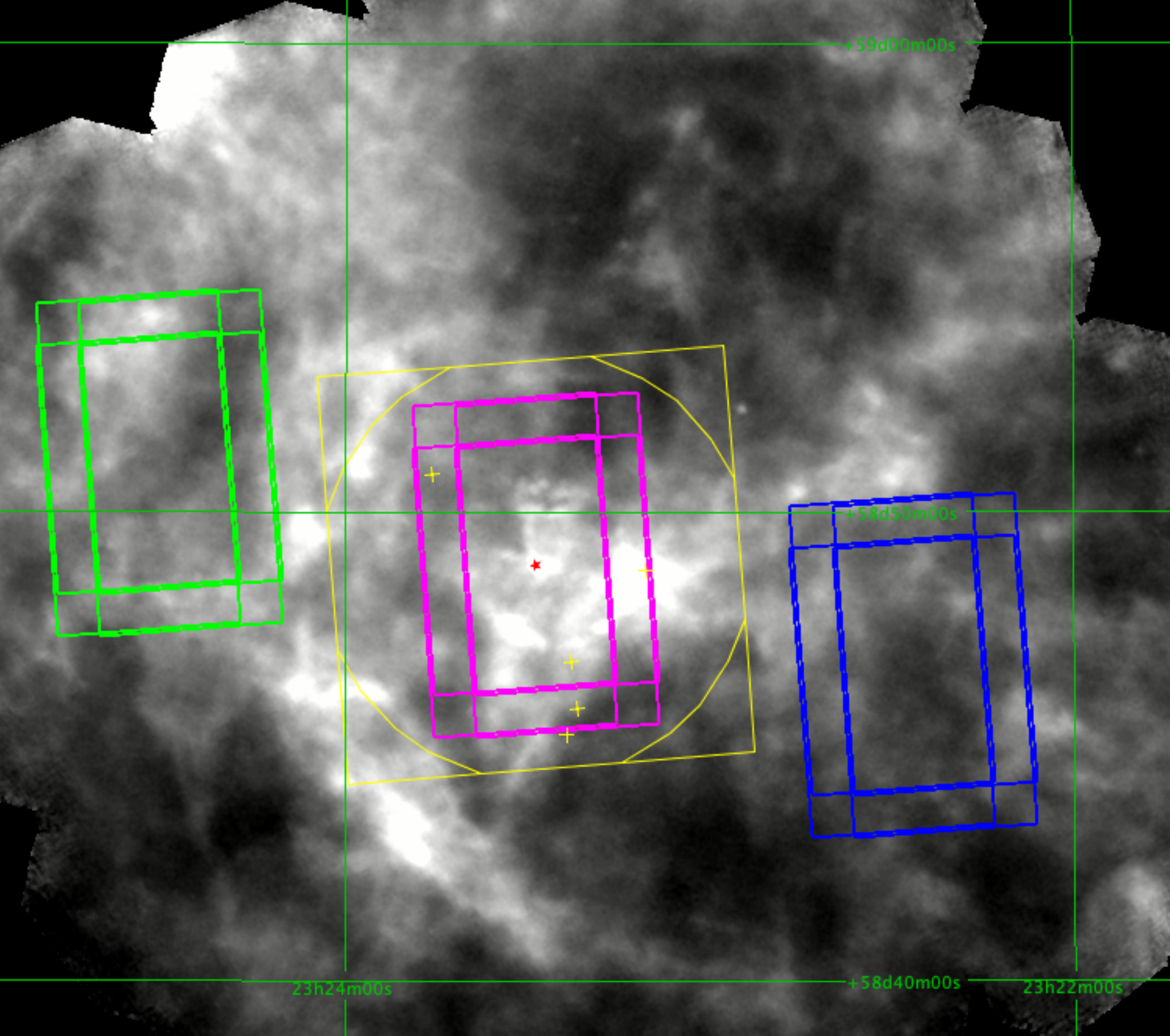}
\caption{The footprint of the SOFIA HAWC+ observation. The target position of Cas~A is marked as a red star. The field of views (FOVs; with dithers) of the target (in purple) and chop and nod (in green and blue; chop and nod FOVs were alternated) are superposed on a \herschel\ PACS 160 \mic\ image. \new{Since the edges of the FOVs (purple squares) of the target had shorter exposures from the dithers and were trimmed, the mosaiced image ended up with a circular shape in Figs. \ref{fig:casaiqu} and \ref{hawcppaccomp}.}
The FOVs of the Focal Plane Imager (FPI) and guide stars (crosses) are marked in yellow.}  
\label{hawcpfootprint}
\end{figure*}

\begin{figure*}
\begin{center}
\hspace{-3.0 mm}
$\begin{array}{cc}
\includegraphics[width=8.4truecm]{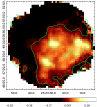} &
\includegraphics[width=8.4truecm]{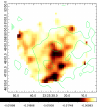} \\
\includegraphics[width=8.4truecm]{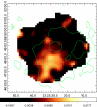} &
\includegraphics[angle=0,width=9.2truecm,height=9.5truecm]{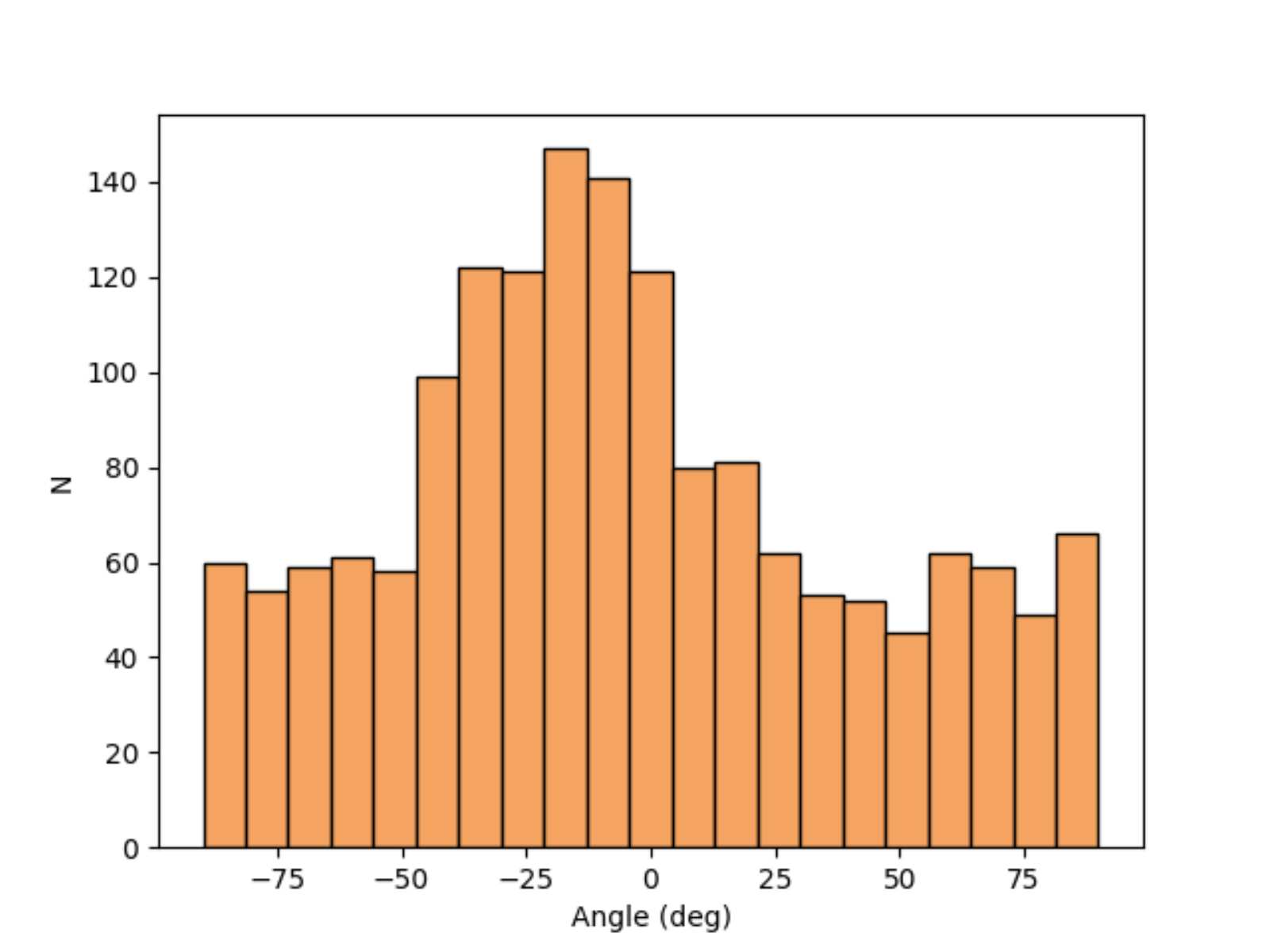}\\
\end{array}$
\caption{SOFIA HAWC+ images of the Stokes I (top left), Q (top right), and U
(bottom right) vectors at Band D (154 \mic) after the images were re-gridded by a
factor of 3$\times$3. The scale bar is at the bottom of each image (Jy
pixel$^{-1}$, where the pixel size is 8.25$''$.). The I, Q, and U maps of
the SNR range 0.020--0.250, -0.005 -- -0.0160, and 0.002 -- 0.015 Jy
pixel$^{-1}$, respectively. The contours from the intensity map are 0.009,
and 0.126 Jy\,pixel$^{-1}$. Note that the edges of the maps show
artifacts.
The distribution of Position Angles (bottom right) after rotation by 90$^{\circ}$ using the HAWC+ polarization angle map that was rebinned by a factor of 3.
}
\label{fig:casaiqu}
\end{center}
\end{figure*}

The SNR Cas~A provides an excellent source for testing theoretical predictions of dust formation and destruction in ccSNRs: it is close enough to allow for a detailed view of gas and dust in the SN ejecta and its associated magnetic field structures, as well as being young enough (${\sim}350\,$yrs) to be ejecta-dominated. The progenitor of Cas~A is believed to be a Wolf-Rayet star with progenitor mass of $15-25\, \text{M}_{\odot}$ \citep{young06}.  The mass of dust predicted to form in the aftermath of the SN explosion is then 0.5 -- 1 M$_{\odot}$ \citep{todini01}. Less clear is how much of any newly formed dust will be ultimately destroyed, with theoretical predictions ranging from 10 -- 90\,per\,cent \citep{bianchi07,bocchio16,kirchschlager19,priestley22}. 

While sub-millimeter (submm) observations of Cas~A seemed to provide the first evidence for large ($\sim$2 M$_\odot$) quantities of colder dust manufactured in the SN explosion \citep{dunne03}, CO observations towards Cas~A suggested that some or all of the submm emission may originate from dust in a foreground molecular cloud complex rather than the remnant \citep{krause04}. Polarization observations taken with the SCUBA camera on the JCMT at 850\,$\mu$m were critical to demonstrate that a large fraction of the submm emission does, in fact, originate from the SNR itself \citep{dunne09}. The fractional polarization of dust in Cas~A was observed to be as high as $\sim$30\,per\,cent \citep[see Fig.\,3 of][]{dunne09}, indicating a highly efficient alignment mechanism for the grains given the short timescale since the explosion. In comparison, the average dust polarization fractions observed in typical interstellar regions are of the order of $2-7\,$per\,cent \citep{curran07}. Assuming that only the highly polarized flux is associated with the SN remnant, then the polarized fraction of the flux indicates a significant dust mass in Cas~A ($\sim$1 M$_\odot$, \citealp{dunne03}). 
However, the SCUBA polarimeter observation at 850 \mic\ mapped only one-third of the total area of
Cas~A \citep[toward the northern and western shells, see Fig. 1 of][]{dunne09}, and the 850 $\mu$m emission includes a significant ($\sim$ 2/3) amount of synchrotron emission.

\begin{figure*}
\includegraphics[scale=0.6,angle=0,width=17.8truecm]{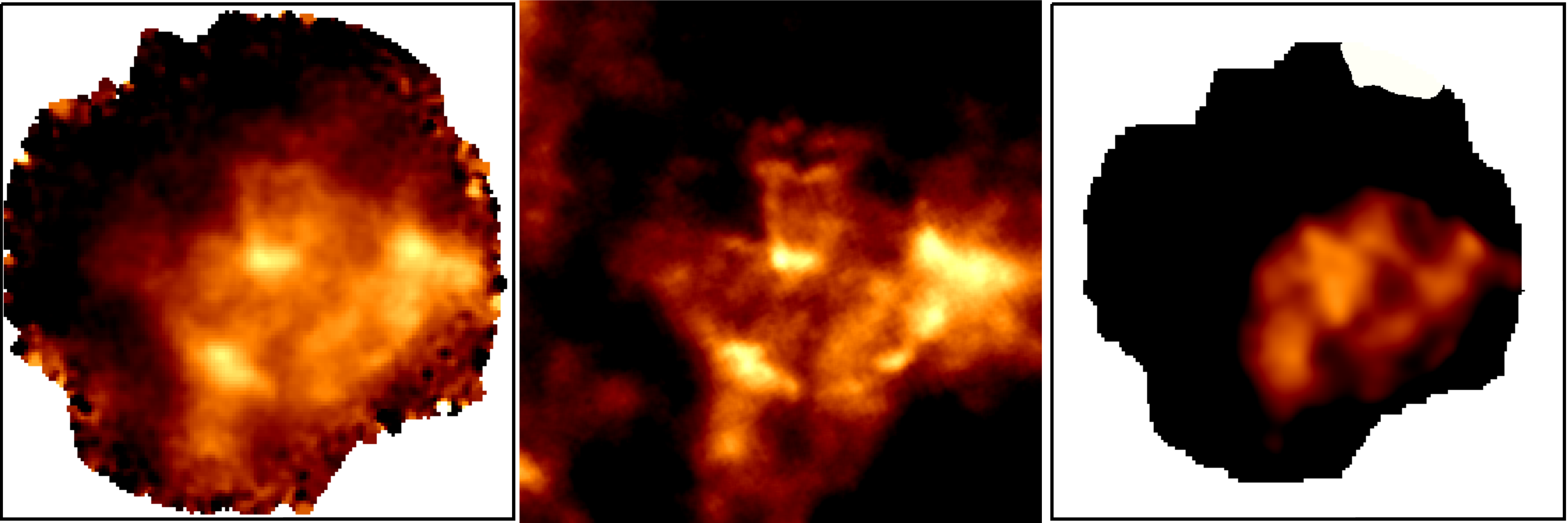}
\caption{The 154 \mic\ HAWC+ observation image in full resolution (left),  \herschel\ PACS 160 \mic\ image (middle), and the ratio map (right) between HAWC+ and Herschel PACS 160 \mic\ (Table \ref{Tphotometry}). The HAWC+ and 160 \mic\ PACS image resolution is 13.6$''$ and 12$''$, respectively. The ratio map is produced after the two maps were smoothed to 13.6$''$, and the background is subtracted from the PACS image. The scale of the ratio map ranges from 0.5  to 1.
The image is centered on R.A.\ $23^{\rm h} 23^{\rm m} 28.52^{\rm s}$ and Dec.\ $+58^\circ$48$^{\prime}51.78^{\prime \prime}$ (J2000) with a FOV of 6.67$'$ $\times$ 6.70$'$.}
\label{hawcppaccomp}
\end{figure*}

\begin{figure*}
\includegraphics[scale=0.6,angle=0,width=16.4truecm]{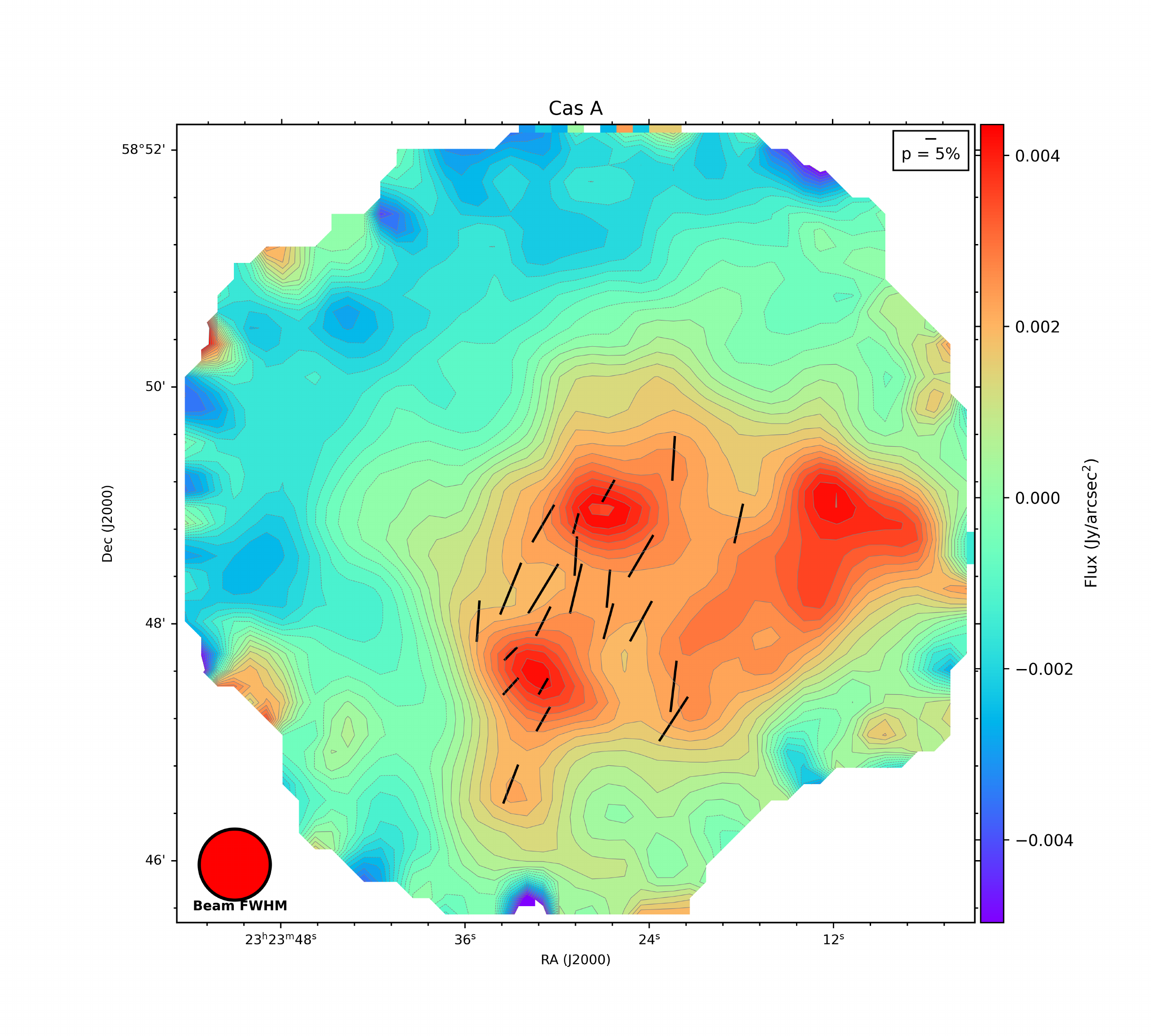}
\caption{
HAWC+ image of Cas~A at 154 \mic\  within a FOV of $\sim 5'\times5'$; the colour scale represents the
total surface brightness. Polarization vectors that fulfill the criteria of $p/\sigma_p > 3$ and $I/\sigma_I$  $>$20 have been rotated by 90$^{\circ}$ to show the inferred magnetic field morphology (black lines). 
Both intensity and polarization maps are binned by a factor of 3 to a spatial resolution of 41$''$.}
 \label{Bpolmap}
\end{figure*}

\begin{figure}
\includegraphics[scale=0.6,angle=0,width=8.truecm,height=8.truecm]{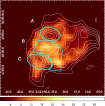}
\includegraphics[scale=0.6,angle=0,width=8.truecm,height=8.truecm]{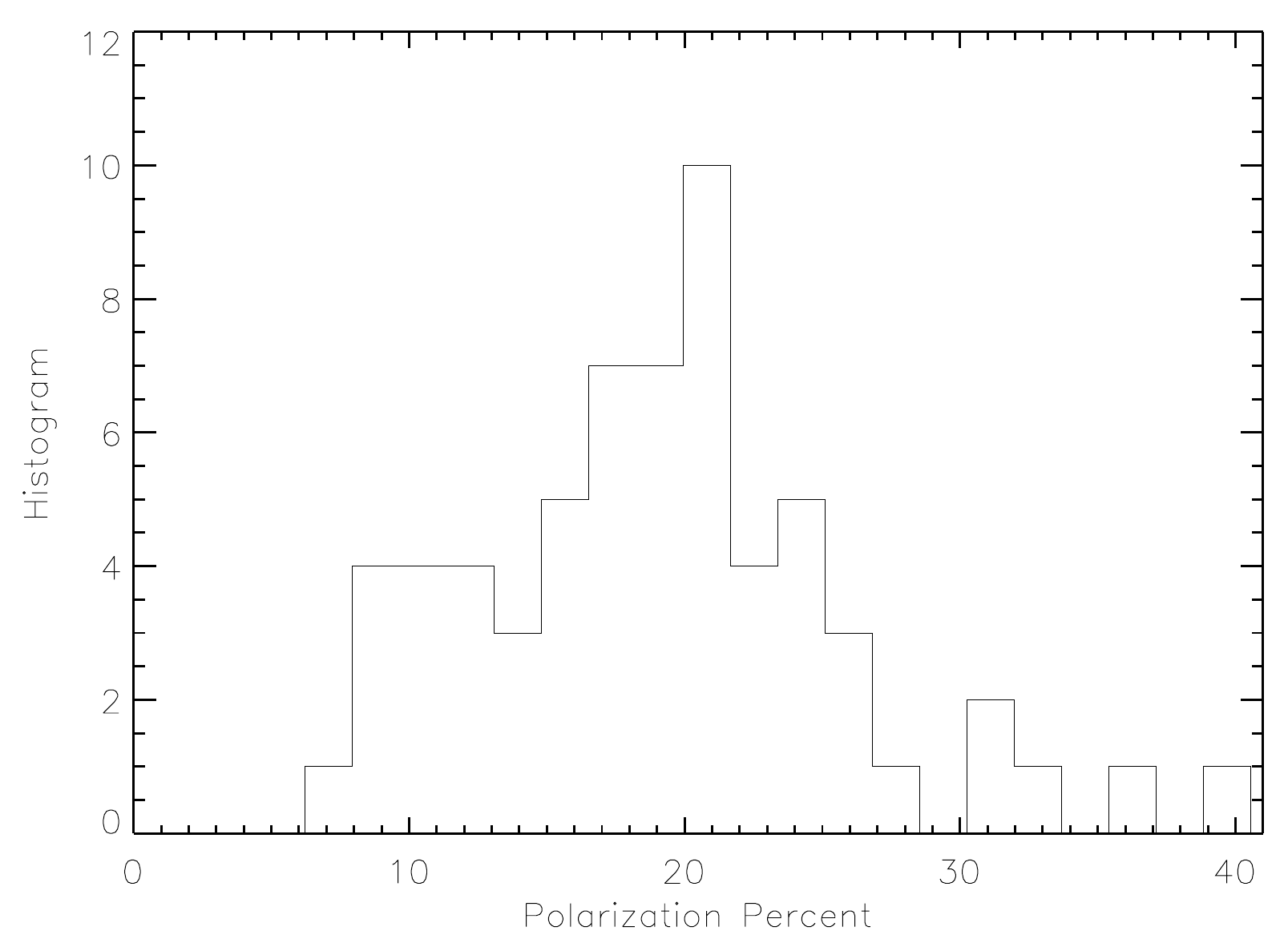}
\caption{\textit{Top:} The polarization percentage map derived from the HAWC+ observations with the 154\mic\ intensity contours overlaid (0.004, 0.0066, 0.009, 0.012, 0.015, and 0.02 Jy/pixel where the pixel size is 8.25$''\times8.25''$).  Marked in blue are the three regions A, B and C corresponding to center, Diffuse, and SE shell, respectively (Table~\ref{Tpol}). These regions were selected to indicate areas with similar dust and $B$-field properties.
\textit{Bottom:} The polarization percent for the 63 elements which meet the S/N criteria of $I/\sigma_I > 20$ and $p/\sigma_p > 3$.   The average polarization fraction is 19.4$\pm$6.7\,per\,cent.}
\label{polpercentregions}
\end{figure}

In this paper, we present polarization observations of Cas~A at 154 \mic\ using the HAWC+ instrument on SOFIA where we are less affected by synchrotron contamination. We investigate the grain alignment, magnetic field strength, and dust properties. We discuss the implication of the dust mass associated with SN-ejecta from the polarization measurements. In an accompanying paper, we present the HAWC+ observation of the Crab Nebula \citep{chastenet22}.

\begin{figure*}
\includegraphics[scale=0.6,angle=0,width=15.4truecm]{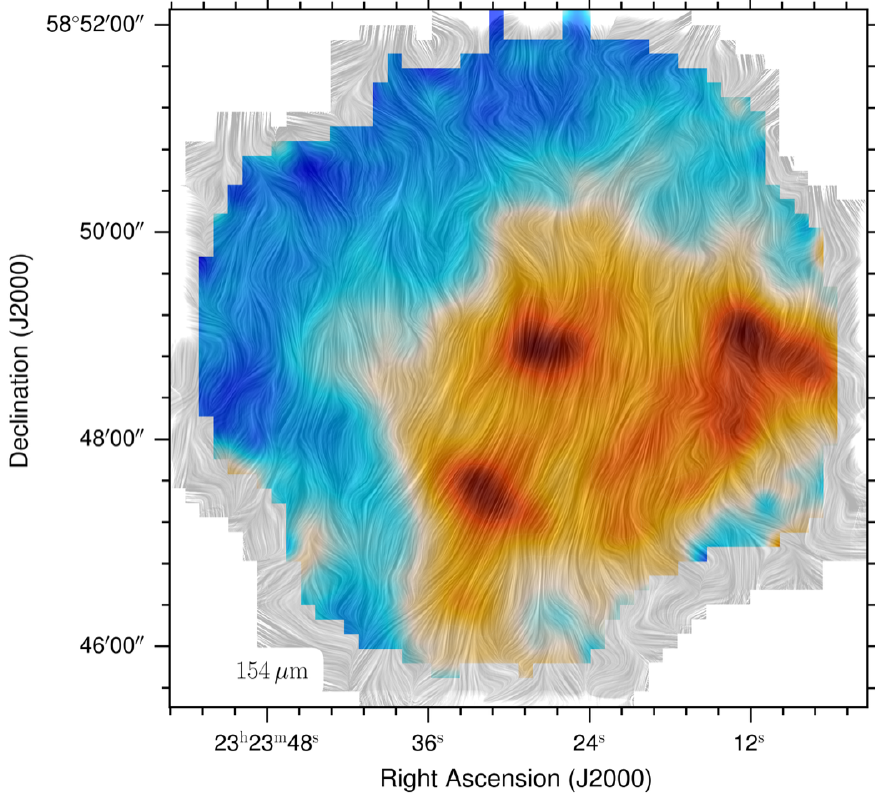}
\caption{Magnetic field direction and strength in Cas~A as shown by the line integral convolve (LIC) map \citep{cabral93} based on the HAWC+ measurements (Fig. \ref{Bpolmap}). The colour scale shows the 154 \mic\ HAWC+ total intensity image (the colours are brown, brownish yellow, blue, and grey, downward in the order of the intensity).}
\label{casapolLIC}
\end{figure*}

\section{Observations}
\label{Sobservations}

\subsection{SOFIA HAWC+ Observations}
We observed Cas~A with HAWC+ \citep{harper18} at band D (154\,\mic) on board SOFIA (AOR\_ID 07\_0047). The exposure time on source is 1.51\,hr (320 sec exposure per cycle $\times$ 17 repeat) and 4.24\,hr including the overhead. F607 on 20190907 (p7504),  F608 on 20190910 (p7506), F610 on 20191008 (p7530),  F620 on 20191009 (p7524), and F621 (p7526) on 20191010. The angular resolution is 13.6$''$ full-width-half-maximum (FWHM) at full resolution. The data were taken using chop-to-nod (C2N) with nod-match-chop (NMC), and the ABBA pattern, with 500$''$ chop throw, 75$^{\circ}$ chop angle, and 4-point dithers. The C2N method was the only available mode at the time of our data acquisition of Cas~A. The observational footprint is shown in Figure~\ref{hawcpfootprint}.

We use the HAWC+ ‘Level 4’ data which corresponds to fully calibrated data combined from different
observing nights. The Level~4 data products are fits files, each containing 19 extensions, including the Stokes $I$, $Q$, $U$, fractional polarization ($p$), position angle of polarization ($\theta$), and polarized flux ($I_\text{p}$), as well as their associated uncertainties, where 
\begin{align}
   \theta = \frac{1}{2}~ {\rm arctan} \frac{U}{Q}.  
\end{align}

The Stokes $I$, $Q$, and $U$ maps are shown in Figure~\ref{fig:casaiqu}.  (For a detailed description of the intensity $I$ and its error $\sigma_{I_p}$, the debiased polarization $p$ and its uncertainty $\sigma_p$, and polarization angles as a function of the Stokes parameters $Q$ and $U$, we refer the reader to \citealt{gordon18} and the SOFIA handbook{\footnote{\url{https://www.sofia.usra.edu/sites/default/files/Instruments/HAWC_PLUS/Documents/hawc_data_handbook.pdf}}}.)

The position angle (PA) provided by the pipeline product shows the polarization direction and not the direction of the magnetic field.
To investigate the inferred magnetic field morphology and angles, the PA is rotated by 90$^{\circ}$, and the result is shown in Figure \ref{fig:casaiqu} (bottom right).
The histogram of PAs reveals that the PAs peak at -20$^{\circ}$ and the number of pixels with PAs between -50$^{\circ}$ and 25$^{\circ}$ is larger compared to that of the background.

Final data products are delivered with a pixel scale equal to the Nyquist sampling (4.68$''$ at 154 \mic). Since the polarization map at full resolution did not have sufficient signal-to-noise,
we resampled the original image by $\sim$3$\times$3 (from 146$\times$146 pixels to 49$\times$49 pixels) using a $`$congrid' python routine. We use the option of nearest $`$neighbor' to interpret the data with the closest value from the original data.
The re-gridded image has 3$\times$3 super-pixels, equivalent to a beam size of 41$''$.

\subsection{ \spitzer\ spectroscopy and \herschel\ Far-IR Images} 

We used archival \spitzer\ mid-IR spectroscopy and \herschel\ far-IR images of Cas~A. The observations of Infrared Spectrograph (IRS) mapping in low-resolution (low-res, hereafter) covered the entire SNR Cas~A and their detailed descriptions were presented by \cite{ennis06}, \cite{rho08}, \cite{smith09}, and \cite{deLaney10}.
The IRS high-resolution (high-res) mapping covered limited area (center, NE, SE, SW) of Cas~A as described by \cite{isensee10} and \cite{isensee12}. The resolution of high-res spectra (R$\sim$600) is a factor of 6 higher than that of low-res spectra.  

\herschel\ images were taken at 70, 100, 160, 250, 350, and 500\,\mic, and their spatial resolutions correspond to 
6, 8, 12, 18.1, 24.9, and 36.4$''$, respectively.  Full details can be found in \cite{barlow10} and \cite{deLooze17}.

\subsection{Comparison of SOFIA/HAWC+ and \herschel}

As a sanity check, we compare the SOFIA HAWC+ 154 \mic\ map with the \herschel\ 160 \mic\ PACS image of Cas~A in Figure~\ref{hawcppaccomp}. The morphology of the two maps is similar to each other, though the resolution (13.6$''$ and 12$''$ respectively) 
and the bandwidths (34 and $\sim$80 \mic\ respectively) differ somewhat. To compare the flux calibration between the two images, a ratio map (Fig.\,\ref{hawcppaccomp}) was made after smoothing the images to the same resolution. To be consistent with the HAWC+ observations (which have already had a "background subtraction" via its chop-nod process) a background level was removed from the \herschel\ map. This was carried out using two background regions approximately at the chop-nod-positions (Fig. \ref{hawcpfootprint}) centered on (R.A.,\,Dec.,\,$a$,\,$b$,\,PA) = (23:22:25.3419,+58:45:42.987, 269$\arcsec$, 370$\arcsec$, 2.6$^{\circ}$) and (23:24:31.2387,+58:51:18.571, 269$\arcsec$, 370$\arcsec$, 2.6$^{\circ}$) where $a$ and $b$ are the minor and major axes of the box. The final ratio map has individual pixel ratios ranging from 0.7 -- 1.2 with a total flux ratio estimated for Cas~A of $F_{154}/F_{160}$ of 0.85$\pm$0.27. 

\section{Results}
\subsection{Polarization Detection in Cas~A}

The HAWC+ $B$-field polarization vectors of Cas~A are overlaid on the HAWC+ 154 \mic\ intensity map in Figure \ref{Bpolmap}. Both the polarization vectors and the intensity map have been rebinned as described earlier. Here we show only the high signal-to-noise pixels (defined here as $p/\sigma_p >3$ and $I/\sigma_I >20$), resulting in a total of 63 polarization "elements". The average polarization fraction in these elements is 19.4 $\pm$ 6.7\,per\,cent (Table \ref{Tpol}), though the polarization fraction across the SNR varies from 5 - 30 per cent (see Figs. \ref{Bpolmap} and \ref{polpercentregions}). As first observed in \citet{dunne09}, we therefore have direct evidence that the grains in Cas~A are elongated since spherical grains cannot emit polarized radiation \citep[see][and references therein]{lazarian15, hoang18, kirchschlager19a}. Our detection of polarized emission shows that the grains are elongated \citep{hildebrand95, draine21b} and aligned with a preferred direction in space
\citep[][see Section~\ref{sec:alignment2} for details]{Hoang.2022}.

Surprisingly, we see strikingly strong polarization from dust grains in the center and southeastern shell of the SNR in the HAWC+ data, in contrast to the 850\mic\ observations where the highest polarization fractions were observed in the northern shell \citep{dunne09}.

A map of the polarization $p$ is shown in Figure \ref{polpercentregions}. There are three primary areas in which polarization is detected. (i) The center of the SNR (labelled region `A' in Fig.~\ref{polpercentregions}, Table~\ref{Tpol}) associated with unshocked material such as [Si~II]  \citep{smith09, rho08, barlow10}; (ii) region `B' extending between the two bright 154 \mic\ regions in the center and south-eastern shell in which the polarization signal is anticorrelated with the 154 \mic\ emission; and (iii) the bright 154 \mic\ emission region on the southeastern shell (Region `C').  The lack of significant polarization signal to the west is likely due to contamination in this region from interstellar emission in the line of sight \citep{deLooze17}. Figs. \ref{Bpolmap} and \ref{polpercentregions} suggest an anti-correlation between the intensity ($I$) and polarization percent ($p$) in that the polarization is weaker when the 154 \mic\ intensity is brighter. Characterization of the $p-I$ anti-correlation is described further in Section \ref{Spolintensity}.

\begin{figure*}
\includegraphics[scale=0.6,angle=0,width=7.2truecm]{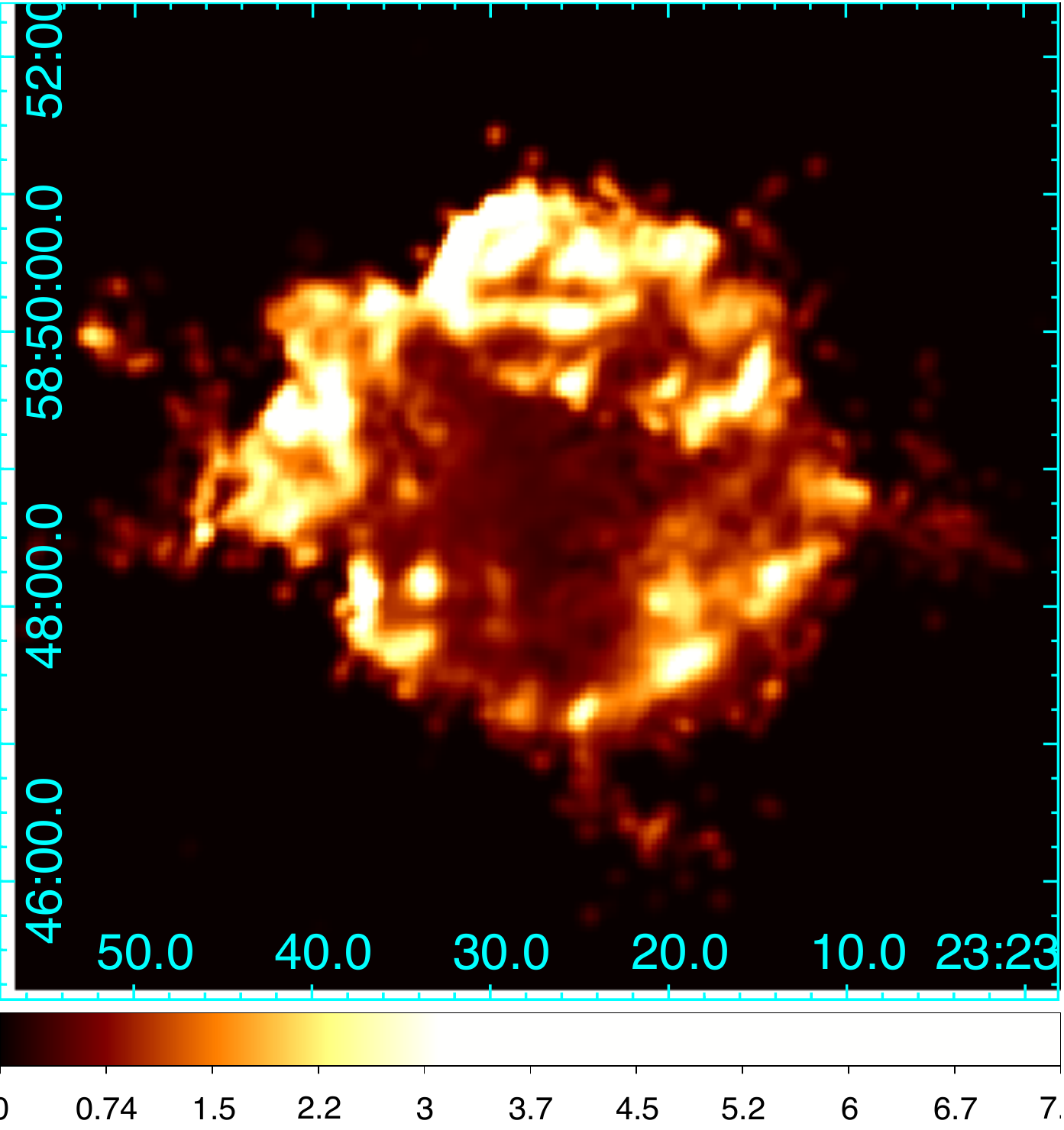}
\includegraphics[scale=0.6,angle=0,width=7.2truecm]{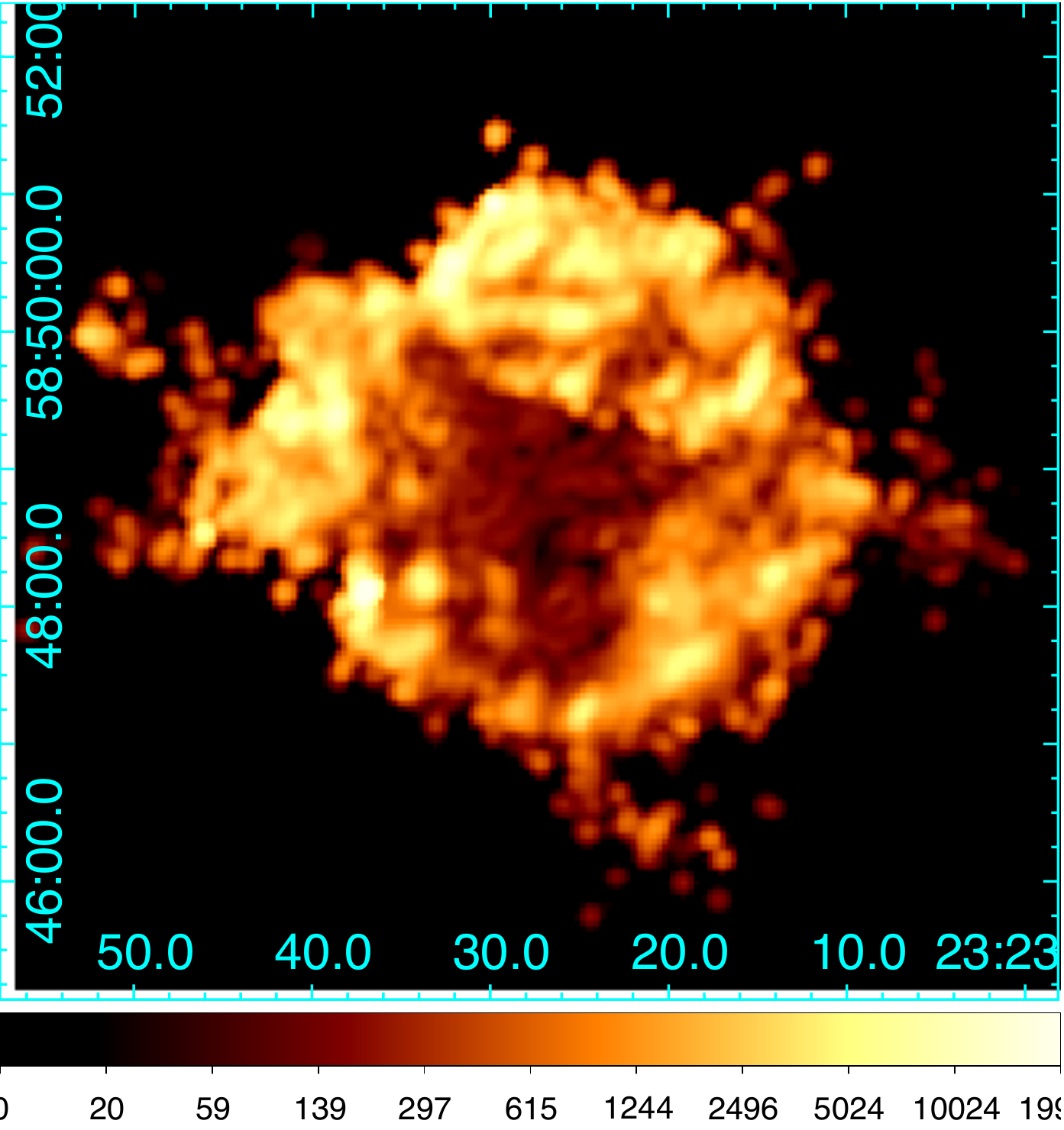}
\caption{(\textit{Left}): Ratio of the  18\,\mic\ to 33\,\mic\ maps from \spitzer\ of \siiif. 
 \textit{(Right):} 
 Density map derived from the ratio map. The densities at the ejecta shell region are between 10$^{4}$ and 10$^{5}$ cm$^{-3}$ and the
interior has a density of a few 100\,cm$^{-3}$. The average numbers for given regions
are listed in Table \ref{Tpol}.}
\label{densitymap}
\end{figure*}

\begin{figure}
\includegraphics[scale=0.6,angle=0,width=8truecm]{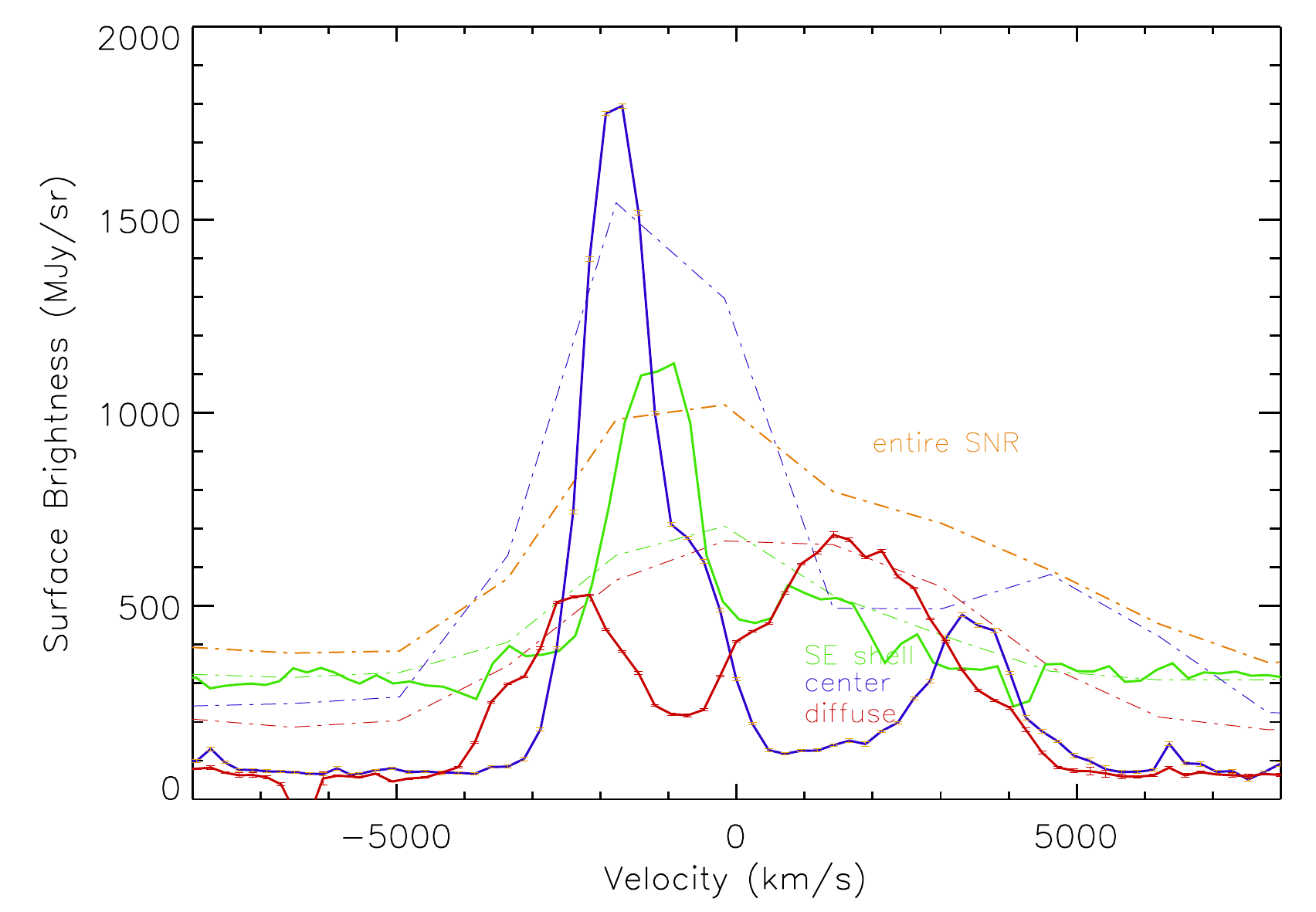} 
\caption{The velocity profiles of \siliif\ from \spitzer\ centered at 34.8152 \mic\ for the four regions listed in Table~\ref{Tpol}. Regions A-C (Fig.~\ref{polpercentregions}) are indicated by the blue, red and green colours with the entire SNR in yellow. High and low-res spectra are shown with solid and dashed-dotted lines, respectively.\label{velocitydispersion} }
\end{figure}

A line integral convolution \citep{cabral93} plot is shown in Figure~\ref{casapolLIC} to show the overall large-scale magnetic field morphology from the HAWC+ polarization observations. The polarization direction in Cas~A is north to south (PA ranges from -50$^{\circ}$ to 25$^{\circ}$ peaking at -20$^{\circ}$ where the positive angle is measured from N-E). The average PA is -20.24$^{\circ}$. The observed PAs from of the dust emission are not particularly radially aligned in contrast to the synchrotron radiation observed in the radio \citep{dunne09}.

\begin{figure}
\includegraphics[scale=0.6,angle=0,width=8truecm]{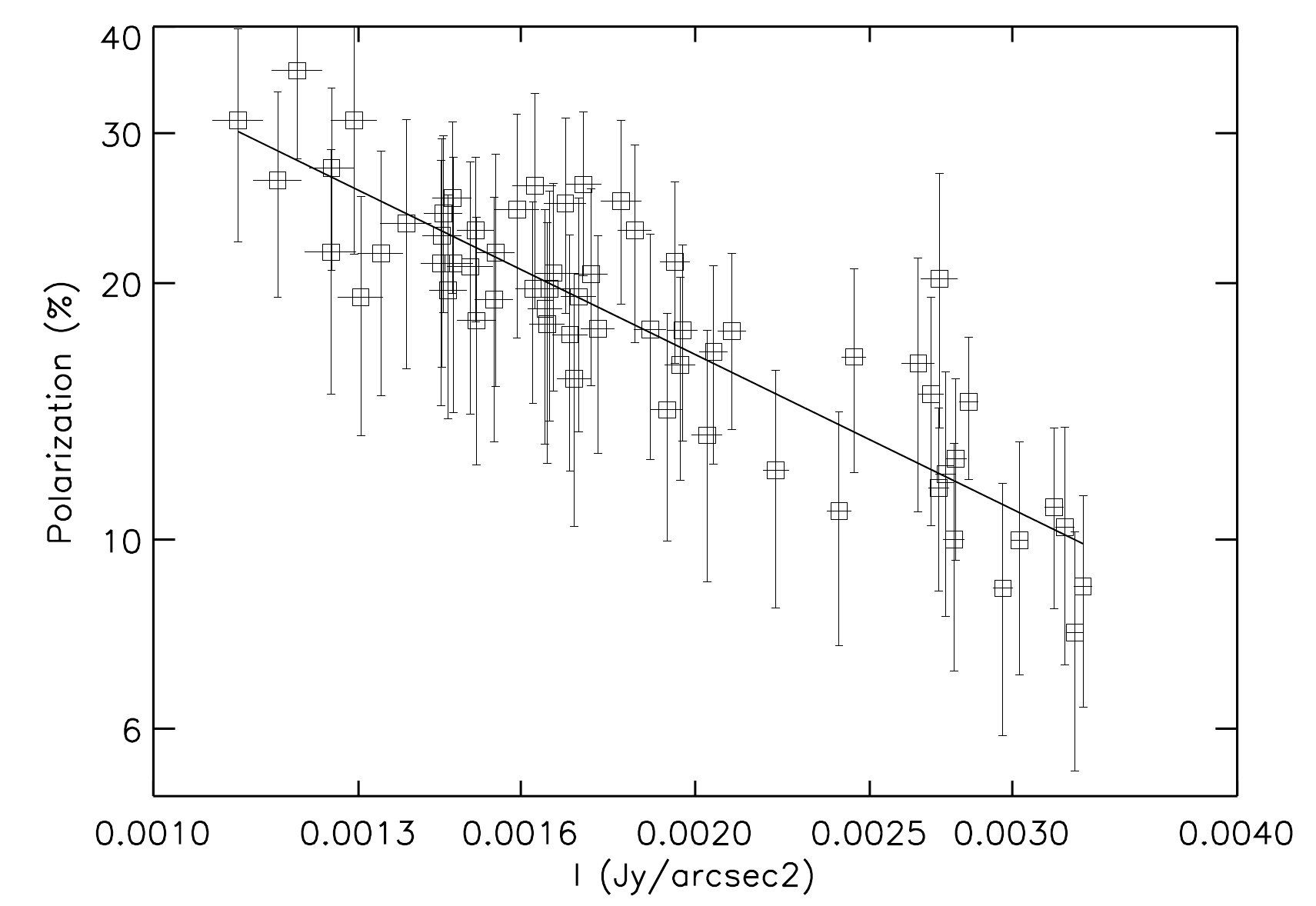} 
\caption{The polarization $p$ fraction as  a function of the total intensity $I$ of the 63 polarization vectors.
A single power-law fit $p \propto I^{\alpha}$ is shown (solid line) with slope  $\alpha$ = -1.03$\pm$0.06.\label{polintrelation} }
\end{figure}

\begin{figure*}
\includegraphics[scale=0.6,angle=0,width=14.0truecm]{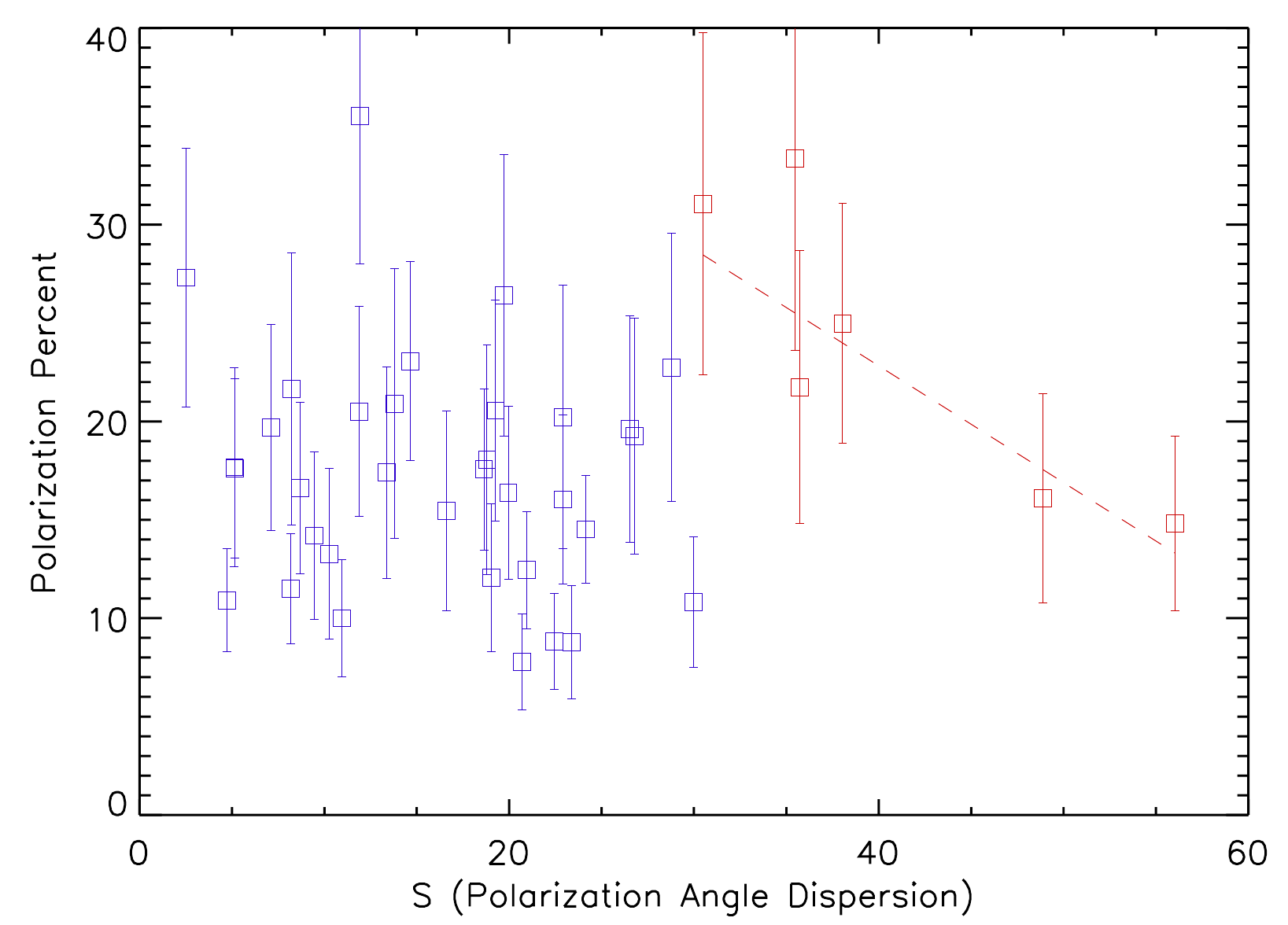} 
\caption{Variation of the polarization fraction on the absolute value of polarization angle dispersion ($S$) with a linear fit for $S\,>$$\sim$30 (solid line) with the representative errors of the polarization percent. The polarization fraction shows a  dependence on the angle dispersion ($S$) for $S\,>$$\sim$30 as we expect the magnetic field becomes more tangled with a higher dispersion $S$. The definition of $S$ is given in the text (Section \ref{Sdepol}).}
\label{polparam}
\end{figure*}
\subsection{Magnetic Field}

We derive magnetic field strengths using the Davis-Chandrasekhar-Fermi (DCF) method \citep{davis51, chandrasekhar53,falceta-Goncalves08,li21} based on the observed velocity dispersion, gas density, and the dispersion of polarization angles. We used a \spitzer\ IRS cube to derive the density and velocity dispersion using two \siiif\ lines at 18.7 and 33.6 \mic\ to create a density map \citep{rho08, smith09}. We extracted the \siliif\ 34.8 \mic\ line using \spitzer\ IRS data from the regions of interest in Figure~\ref{polpercentregions} to derive the velocity dispersion (Fig.~\ref{velocitydispersion}) and use the HAWC+ polarization map to derive the dispersion in polarization angle $\theta_p$.
The resulting parameters are summarized in Table \ref{Tpol}.

\subsubsection{Volume density}

We derive the density map in Figure \ref{densitymap} from the \siiif\ line ratio of $18.7$ and $33.5$ \mic\ lines assuming a temperature of 6000 K \citep{smith09}. The density map indicates that the central blob and SE shell where we see low polarization (${\sim} 6-8$\,per\,cent) has a density of $100-200\,$cm$^{-3}$ and $500-10^4\,$cm$^{-3}$, respectively. In contrast, the region between these bright dust regions (fainter in the 154 \mic\ map, see Region `B' in Fig.~\ref{polpercentregions}, top) has a gas density of $\sim$500 cm$^{-3}$ and a polarization of 15 \,per\,cent. 
One would predict that higher gas densities can lead to higher collision rates between the dust and gas, which causes stronger randomization. This decreases the alignment efficiency, resulting in reduced polarization \citep{hoang21}.

\subsubsection{Velocity Dispersion}
\label{Svelocity}

We use the \siliif\ line at 34.8152 \mic\ from \spitzer\ IRS spectral mapping data \citep{ennis06,deLaney10, rho08, smith09}, since this line has a higher signal-to-noise than the \siiif\ lines. We extracted the spectra from the central, unshocked region (Region A), the southeast shell (Region C), and the diffuse region (Region B) between the two (Fig.~\ref{polpercentregions}, top) using both high-res and low-res spectra. The spectra from the entire HAWC+ map are using low-res spectra only because the entire SNR was not covered with high-res spectra \citep{isensee10,isensee12}. The velocity line profiles are shown in Figure \ref{velocitydispersion}. We fit the \siliif\ line using multiple components of the Gaussian velocity profile and the velocity dispersion is estimated with $\sigma_V = \text{FWHM}/2.355$ \citep{hwang21}. The derived velocity properties are summarized in Tables \ref{Tpol} and \ref{Tvelfwhm}.

The velocity dispersion using the high-res spectra is between 450 and 750 \kms. The Doppler or expansion motion is measured using the shifted velocity component ($V_{\rm shift}$,  Table \ref{Tvelfwhm}). However, since the SN ejecta is knotty with small-scale structures as seen in the HST images \citep{fesen06}, the velocity profiles of the large regions may include numerous knots. Therefore, the derived velocity dispersion $\sigma_V$ may be an upper limit. Spectra with higher spatial ($0.6-1''$) and spectral resolution ($R\sim2700$; \citealt{gerardy02,lee12}) were taken toward different parts of Cas~A from the regions we need. \new{Their position-velocity maps show many knots with a velocity range of 400 - 1000 \kms}, indicating the velocity dispersion may be comparable to the numbers using high-res spectra in Table \ref{Tpol}, but some of the knots show a lower velocity dispersion of $\sim$200 \kms. Thus the velocity dispersion using higher resolution spectra is likely a factor of a few smaller than those in Table \ref{Tpol}.

\subsubsection{Polarization angle dispersion}
\label{Sangledispersion}

We estimated the dispersion in polarization angles using  polarization measurements and their errors for each pixel.
The uncertainties of individual PA values are important parameters when calculating the dispersion as described below.

Taking into account measurement errors in our polarization data, we estimate  $\sigma_\theta$ as the weighted standard deviation of polarization angles ($\theta$ = PA) \citep{li21}:
\begin{equation}
{\sigma_\theta^2  } = \frac{N}{N-1}\,\frac{1}{w}\,
\sum\limits_{i=1}^N w_i \, (\theta_i - \bar{\theta}_w)^2 ,
\label{eq:sigmatheta}
\end{equation}
\noindent
where $N$ is number of independent measurements (63), $w_i = 1/\sigma_i^2 $ the weight of measurement $i$ given the measurement error in PA $\sigma_i$, $w = \sum\limits_{i=1}^N w_i$, and $\bar{\theta}_w = (1/w)\, \sum\limits_{i=1}^N w_i\, \theta_i$ is the weighted mean polarization angle. The average polarization angle $\theta$ is -20.2$\pm$18.2$^{\circ}$, and the median value of the dispersion $\sigma$ is 7.7$\pm$1.0$^{\circ}$. The average angle may infer the overall flow of the magnetic field direction as shown in Figure \ref{casapolLIC}. The weighted mean $\theta_w$ is -20.37$^{\circ}$. 
Estimated from the Equation \ref{eq:sigmatheta},  the dispersion of the polarization angle, $\sigma_\theta$ is 17.2$\pm$2.2$^{\circ}$. 

The histogram of PAs for the 63 polarization elements shows a similar profile to Figure \ref{fig:casaiqu} (bottom right), but with a smaller number of pixels.
We can fit the histogram using the three-component Gaussian model as we have done for the velocity dispersion. We find that the strongest component has -23$\pm$7.8$^{\circ}$. The second and third components are -4.4$\pm$4.6$^{\circ}$ and -42$\pm$1.9$^{\circ}$, respectively. 

\subsubsection{Strength of Magnetic Field}
\label{sec321}

The mean strength of the plane-of-sky magnetic field \citep[][]{li22,falceta-Goncalves08}
is given by: 
$$B_0 = \frac{(4\pi\rho_{\text{gas}})^{1/2}\sigma_V}{\tan\sigma_\theta}\\
= 0.383 \sqrt{n_{\text{gas}}}\, \frac {\sigma_V   }{\tan\sigma_{\theta}}  ~~~\mu {\rm G},
$$
where $\rho_{\text{gas}}$ is the gas mass density (g\,cm$^{-3}$),
$n_{\text{gas}}$ is the gas number density in cm$^{-3}$, $\sigma_V$ is the measured velocity dispersion in \kms, and $\sigma_{\theta}$ is the dispersion of the magnetic field orientation.
The derivation of this equation and the related DCF method are summarized in Appendix A of \citet{li21}.
We estimate the magnetic field based on these equations using the input parameters in Table \ref{Tpol}. The magnetic field strength for the HAWC+ observed regions amounts to 9 -- 40 mG (Table \ref{Tpol}). 
The maximum magnetic field is observed in the shocked SE shell (39 mG), whereas the central un-shocked region shows a lower field strength of 9 mG. It is surprising to find such a strong magnetic field, although the shocks increase the magnetic field by the second-order Fermi acceleration  \citep{cowsik84,rho03}. When we use a factor of 4 smaller velocity dispersion (see Section \ref{Svelocity}), the magnetic field strength ranges from 2 -- 10 mG. Note that there are various uncertainties in estimating the magnetic field strength using the DCF method, but it is the only way we know to estimate the field strength with polarization measurements. There have been many magnetic field strength estimates made for different regions of Cas~A. We compare our measurements with previous values and discuss them in Section \ref{Smagneticfieldcomp}.

\begin{table*}
\begin{tabular}{llccccccc}
\hline \hline
Region$^a$ & Polarization & Polarization& $T_{\rm d}^b$ & $n_{\rm gas}^c$ &$\sigma_V^d$ &$\sigma_{\theta}$ &  $B$$^e$ \\
     & fraction (\%)  &   angle ($^{\circ}$)& (K) & (${\rm cm}^{-3}$)&(km~s$^{-1}$) & ($^{\circ}$) & (mG)\\ \hline
Entire SNR & 19.4$\pm$6.7 &-20.2$\pm$18.2 & 30 & 686      & 465 (1552) & 17.2$\pm$2.2  & $\sim$4 (15) \\
A. Center (unshocked)& 11.9$\pm$3.8  & -8.1$\pm$8.3 & 28  & 121  & {450} (1369)& 11.2$\pm$5.0 & 2 (9) \\
B. Diffuse &   12.0$\pm$3.0&  -28.9$\pm$8.1 & 33& 673  & 575 (1569) & 8.1$\pm$2.7 & 10 (39)  \\
C. SE Shell &   23.5$\pm$5.7  & -19.2$\pm$7.3  & 29 &501  & 748 (2726) & 18.1$\pm$2.6&  5 (20) \\
\hline \hline
\end{tabular}
\caption{
Polarization properties of Cas~A and the Results of the Davis-Chandrasekhar-Fermi Analysis. 
$^a$The entire SNR is the region covered by the HAWC+ observations. Regions A, B, and C are marked in Figure \ref{polpercentregions} (top). 
$^b$The dust temperature (T$_d$) is from the temperature maps derived by De Looze et al.\, (2017). 
$^c$The density is derived from the [S~III] line ratio of $18.7$ and $33.5$ $\mu$m lines. $^{d}$The velocity dispersion numbers are using [Si~II] lines of high-res and low-res (in parenthesis) spectra (see Fig.~\ref{velocitydispersion} and Table \ref{Tvelfwhm}).
The velocity dispersion using high-res is a factor of $\sim$4 smaller than using low-res spectra. Since no high-res spectra exist for the entire SNR, we use $\sigma_V$ using low-res divided by 4. Typical errors of the dust temperature ($T_{\rm d}$), gas density ($n_{\rm gas}$), and velocity dispersion ($\sigma_V$) are $\sim$5 K, 10 cm$^{-3}$, and 5 \kms, respectively. $^{e}$The magnetic field (B) strength is derived using DCF method and $\sigma_V$ using high-res spectra (low-res spectra in parenthesis). See the text for details.}
\label{Tpol}
 \end{table*}

\begin{table}
\begin{center}
\begin{tabular}{llcc}
\hline \hline
Region & $V_{\rm shift}$ (\kms) & FWHM (\kms) & $\sigma_V$  (\kms) \\
\hline
Centre & -1782$^a$ & 1062 &{\bf 450}\\
       &  -536 & 976 & 414\\
       & 3405  & 1484 & 630\\
       \hline
Diffuse & -2403$^a$ & 1762 & {\bf 748} \\
        & 1632 & 3162 & 1342 \\
        \hline
SE  &-1200$^a$&1354 &{\bf 575}\\
   & 1172 & 1687 & 716\\
\hline\hline
\end{tabular}
\end{center}
\caption{Characteristics of velocity components using \spitzer\ IRS high-res spectra.
$^a$We selected the most substantial velocity component using the low-res spectra (see Fig.~\ref{velocitydispersion}). When the line strengths are comparable to the others, we selected the smallest line widths.}
\label{Tvelfwhm}
\end{table}

\subsection{Polarization fraction vs. emission intensity}
\label{Spolintensity}
The polarization fraction as a function of total  intensity is shown in Figure~\ref{polintrelation}. The polarization fraction decreases as the intensity of the far-IR emission increases. Though the plot shows a significant scattering of the data, we fit it with a power-law function $p \propto I^{\alpha}$. The slope of best-fit model amounts to $\alpha = -1.03 \pm 0.06$. A similar slope has been found for 30~Dor \citep{tram21}. The steep slope of $\sim -1$ reveals that grain alignment is only efficient in the outer layer of the ejecta of Cas~A, whereas the alignment is lost in the inner layer (\citealt{hoang21}). 

However, the $p-I$ anti-correlation is not sufficient to characterize grain alignment properties because the total intensity $I$ depends on several key parameters, including gas column density and dust temperature, $I \sim N_{\rm H} ~B(T_d)$ =~$n_{\rm gas}~l_{\rm gas} ~B(T_{\rm d})$, where $N_{\rm H}$ is the gas column density, $T_{\rm d}$ is the dust temperature, $B(T_{\rm d})$ is the Planck function, $n_{\text{gas}}$ is the gas density, and $l_{\text{gas}}$ is the length of the line of sight of the emitting gas. To better understand grain alignment, we examine correlations of polarization with dust temperature, density, and velocity dispersion, and we find the following results:

$\bullet$ We observe no dependence between polarization percent and temperature, where we use the dust temperature maps derived by \cite{deLooze17}.

$\bullet$ Material with higher velocity dispersion shows higher polarization fractions (see Table \ref{Tpol}).

$\bullet$ 
There is no clear dependence on the polarization fraction and gas density based on the \siiif\ line ratio. However, when comparing the central and diffuse regions (regions A and B respectively), the higher density gas observed in region B has a correspondingly higher polarization (and higher velocity dispersion) than the central region. 
The polarization $p$ depends on n$_H$ (the hydrogen number density, which can be traced by n$_{\rm gas}$). However, since the observed polarization fraction arises from dust grains in the entire column, the dependence of $p$ on $n_{\rm gas}$ may not be observed because of the spatial variation of $n_{\rm gas}$ along the line of sight. 

A detailed discussion on grain alignment and implications for observed polarization is presented in Section \ref{sec:alignment}.

\subsection{Potential Contaminants In the Polarized Signal from the Chop-nod Mode}

We performed similar calculations as \citet{dunne09} to investigate any biases caused by the subtraction of a constant Stokes vector from the chop-nod (off) location to account for the ISM emission. We simulate the effect of the chop-nod on the resulting polarization properties for a target with polarized levels of 2, 5, and 10\,per\,cent (see Fig.~\ref{FigSim}).
Assuming an ISM polarization of 2\,per\,cent \citep{planck15} and a range of polarization angles, we find that the final polarization fraction $p$ of Cas~A, as provided by the HAWC+ Level 4, data could be boosted (by more than 10 -- 15\,per\,cent; $\sim$3.5, 7 and 13\,per\,cent result in 5.5, 11, and 21\,per\,cent, respectively as shown in Fig.~\ref{FigSim}) and that the position angle could be significantly different (typically a $90^\circ$ difference). More details are given in Appendix~\ref{AppSimu}. The level of bias depends on the intrinsic target polarization properties and could be more important in regions with low signal to noise emission. 
This may provide a potential explanation of the lack of polarization seen in the northern shell of Cas A at 154 \mic\ compared to \citet{dunne09}. If the chop-nod has resulted in an over subtraction of the off-ISM here (because the ISM contribution in the North part is relatively low or the northeastern part of the chop-nod off images has relatively brighter emission than elsewhere in the image as shown in Fig.~\ref{hawcpfootprint}), this would artificially make the resulting Stokes I negative (see Fig.~\ref{FigNorth}). The uncertainty in the background level makes it difficult to quantify the effect on our observations; hence we use the measured debiased polarization fraction from the current pipeline and warn the reader of this issue. Appendix~\ref{AppSimu} presents a chop-nod simulation and shows how the observed and the true polarization per\,cent differ under assumed polarization in the ISM.
Potential contamination via synchrotron and the ISM to the polarization signal in Cas~A are discussed further in Section \ref{Sdustmass}.

\section{Discussion}

\subsection{Cas~A with strong FIR polarization and magnetic field strength} 
\label{Smagneticfieldcomp}

The Band D HAWC+ image of Cas~A shows an average polarization fraction of 19.4$\pm$6.7\,per\,cent and ranges from 5 to 30 \,per\,cent across the SNR. The average polarization fractions in typical interstellar dust are of the order of 2--7\,per\,cent \citep{curran07}, with recent \textit{Planck} observations measuring 1--10 \,per\,cent \citep{planck15}. The maximum polarization fraction of thermal dust emission can reach up to about 20\,per\,cent in several large-scale regions of the Galaxy, such as the second ($l$ = 145$^{\circ}$) Galactic quadrant \citep{planck15}. 

Could the high polarization fractions measured by HAWC$+$ be a result of the surrounding regions and not the SNR? \textit{Planck} observations from $30-353\,$GHz (10,000 - 850 \mic) detect primarily synchrotron emission at the location of Cas~A \citep{planckcollXXXI16}, with an excess above the predicted synchrotron flux observed at 850 \mic\ due to thermal dust emission. We have checked the polarization of Cas~A using the shortest \textit{Planck} wavelength (850 \mic) and found weak polarization (${<}1.2\,$per\,cent) toward Cas~A, likely due to the large beam of the Planck image (note that the map resolution is 10$'$, a factor of 46 poorer than that of HAWC+ at 154 \mic).

The magnetic field strength of Cas~A measured in this work using the FIR emission from dust ranges from 9 to 40 mG (or 2 to 10 mG if the velocity dispersion is assumed to be overestimated by a factor of 4, Table~\ref{Tpol}). For comparison, the field strength of the OMC-1 star-forming region range from 0.8 - 26.2 mG \citep{hwang21}, with 0.115 mG observed in LkH$\alpha$ 101, part of the Auriga–California molecular cloud \citep{ngoc21}. The magnetic field strength $B$, inferred from observations of diffuse and molecular clouds via Zeeman splitting follows a relationship with hydrogen number density, $n_{\rm H}$, such that $B \simeq 0.5\,n_{\rm H}^{0.65} \mu$G. For $n_{\rm H} = 10^4$ cm$^{-3}$, we would expect $B \sim$200 (80 -- 500) $\mu$G \citep[see Fig.~1 of][]{crutcher10}. The magnetic field strength of Cas~A is comparable to that of OMC-1 and more than 10 times stronger than seen in clouds.
Potentially, this can be explained as a result of the impact of the reverse shock on the SN gas. In MHD simulations, the magnetic field strength of a few Microgauss can be amplified to a few hundreds to several thousand of Microgauss from such an interaction \citep{kirchschlager23}.

How does the $B$ field strength derived in this work compare with previous measurements of Cas~A? Zeeman splitting of OH lines (tracing dense gas) towards Cas~A showed a weaker magnetic field strength (9 $\mu$G) than observed (11 -- 40 $\mu$G) in the atomic H{\sc i} line \citep{bregman83, heiles86}. 
These low magnetic field strengths are typical in the ISM, indicating that these clouds likely have not been shocked by the shock fronts of Cas A \citep{verschuur69, schwarz86, heiles86}. 

The magnetic field derived from particle acceleration ranges from 0.1 - 2 mG as estimated using X-ray synchrotron emission \citep{helder12}, radio observations \citep{rosenberg70, atoyan00, abeysekara20} and mid-infrared observations \citep[see the discussion in][]{domcek21}. 
Our measurement of 2 -- 10 mG of the $B$ field strength (Table \ref{Tpol}) is therefore 1 -- 5 times higher than the synchrotron measurements (0.1 - 2 mG). 
We should caution that it is difficult to compare the magnetic field probed by polarized dust emission with that obtained from synchtroton emission. Indeed different magnetic field strengths were found at different wavelengths in 30 Dor \citep{tram22}. 
The high field strengths derived here could also be affected by the uncertainties in the DCF method, including the filling factor of dense gas (as the ejecta knots are highly clumpy) and the associated velocity dispersion of the clumpy knots. Here we provide the first estimate of the magnetic field strength in a SNR using the dust polarization.

Higher spatial resolution dust polarization maps may reveal local
variations in the strength of the magnetic field across the SNR.  More samples of far-infrared polarization observations of SNRs and SNe are needed to investigate further. We note that the polarization fraction and possibly the magnetic strength of dust in the FIR is much stronger than that observed in the radio from synchrotron emission in Cas~A
\citep{anderson95}, indicating that the dust is efficiently aligned. We will discuss this further in the next section.

\subsection{Grain alignment mechanisms in Cas~A}\label{sec:alignment}

\subsubsection{Can grains align with the ambient magnetic field?}
\label{sec:alignment1}
We have used dust polarization to infer the morphology and strength of the magnetic field using the DCF method. This assumes that dust grains are aligned with the magnetic field (i.e magnetic alignment). Here we check whether this assumption is valid for Cas~A. 

Following the modern grain alignment theory (e.g., \citealt{andersson15,lazarian15}), the first requirement for magnetic alignment is that grains must have paramagnetic properties. This requirement is satisfied when Fe atoms having unpaired electrons are incorporated in the grains, which is justified for silicate grains or metallic grains. The second requirement is that the Larmor precession of such paramagnetic grains around the magnetic field must be faster than the randomization of the grain's orientation by gas collisions. Following \cite{hoang21}, the Larmor precession time of the grain around the magnetic field is 
\begin{equation}
   t_{\rm Lar}\simeq 0.0042~\hat{\rho}~a_{-5}^2\left( \frac{10^{-4}}{\chi(0)}\right)\left(\frac{10 ~{\rm mG}}{B}\right) ~{\rm yr},\label{eq:tLar}
\end{equation} 
where $\hat{\rho}=\rho/(3~{\rm g cm}^{-3})$ with $\rho$ the grain mass density, $a_{-5}=a/(10^{-5}$cm) with $a$ the grain size, $\chi(0)$ is the magnetic susceptibility of the grains, and $B$ is the magnetic field strength \citep[see also][]{hoang16}.

One can see that the Larmor precession time is rather short, with $t_{\rm Lar}$ = 0.0042 yr for Cas~A assuming $a = 0.1$ \mic\ and $B=10$ mG as listed in Table \ref{Tpol}. 

The characteristic timescale of the grain randomization by gas collisions (i.e., gas damping time) is 
\begin{equation}
    t_{\rm gas}\simeq 830~\hat{\rho}~a_{-5}\left(\frac{1}{n_{3}~T_{\rm gas,3}^{1/2}}\right) {\rm yr},\label{eq:tgas}
\end{equation}
where $n_{3}=n_{\rm H}/10^{3}{\rm cm}^{-3}$
and $T_{\rm gas,3}=T_{\rm gas}/10^{3}{\rm K}$ with $T_{\rm gas}$ the gas temperature (see, e.g., \citealt{hoang21}). The equation implies $t_{\rm gas}\approx 201$ yr for n$_{\rm H} = 686\,$cm$^{-3}$ and T$_{\rm gas} = 6000\,$K (as inferred from the temperature of the \siiif\ line). 

Equations (\ref{eq:tLar}) and (\ref{eq:tgas}) reveal that $t_{\rm Lar}\ll t_{\rm gas}$, i.e., the Larmor precession time is much shorter than the gas damping time of dust grains in Cas~A. 
Therefore, grain alignment is expected to occur with the ambient magnetic field of the SN ejecta, i.e. dust polarization can be a reliable tracer of the magnetic field.

We note that, the lower limit of the magnetic susceptibility $\chi(0)$ obtained from the magnetic alignment condition of $t_{\rm Lar}<t_{\rm gas}$ presents a lower constraint on the Fe abundance locked in the 
silicate lattice. For instance, the numerical prefactor in Equation ~\ref{eq:tLar} is obtained for the paramagnetic susceptibility of $\chi(0)$ = 10$^{-4}$,
which corresponds to about 10\% of cosmic abundance of Fe locked in silicate lattice \citep[see Eq.~(4) in][]{hoang16}. 
By comparing Equations (\ref{eq:tLar}) and (\ref{eq:tgas}) above, a much lower value of $\chi(0)$ also satisfies the criterion of $t_{\rm Lar}$ $<$ $t_{\rm gas}$ for the magnetic alignment.

\subsubsection{Grain alignment by radiative torques (RATs)}
\label{sec:alignment2}

Next we discuss the physical mechanisms that can efficiently align dust grains within the SNR. Grain alignment via radiative torques, RATs \citep{dolginov76,draine97,lazarian_hoang:2007} is thought to be the dominant process for aligning grains with sizes $a > 0.05$ \mic\ \citep{andersson15}. RATs can align elongated grains against gas collisions when the grains rotate at an angular velocity greater than their thermal rotation value, i.e., suprathermally. Moreover, the combination of fast Larmor precession and RAT alignment also result in grains aligned with the magnetic field (the so-called B-RAT mechanism, e.g. \citealp{pattle2019}).

The minimum size of dust grains that can be aligned by RATs depends on the local gas properties and radiation field, as given by \cite{hoang21},
\begin{eqnarray}
    a_{\rm align} &\simeq &0.15~\hat{\rho}^{-1/7}~ \left(\frac{n_{3}~T_{\rm gas,3}}{\gamma_{0.3}~U}\right)^{2/7}
    \times \left(\frac{\bar{\lambda}}{1.2\,\rm \mu m}\right)^{4/7}\nonumber\\
&\times&    \left(\frac{1}{1+F_{\rm IR}}\right)^{-2/7} ~\rm{\mu m},\label{eq:aalign}
\end{eqnarray}
where $\gamma_{0.3}=\gamma/0.3$ with $\gamma$ the anisotropy degree of the interstellar radiation field (ISRF) inside Cas~A, and $\bar{\lambda}$ is the mean wavelength of the ISRF which has a typical value of $\bar{\lambda}=1.2 \mu$m \citep{draine97}, and $U$ is the strength of the radiation field, which is defined by the ratio of the radiation energy density to the typical value of the ISRF in the solar neighborhood \citep{Mathis1983}. $F_{\rm IR}$ is the dimensionless parameter describing the rotational damping by IR emission relative to gas damping. For the dense gas of Cas~A exposed to the typical ISRF of $U=1$, one has $F_{\rm IR}\ll 1$ (see \citealt{hoang21}).

The above equation implies minimum grain sizes of $a_{\rm align}\sim 0.078, 0.15$ and $0.29 \mu$m for $n_{\rm H}=10^{2},10^{3}$ and $10^{4}\, {\rm cm}^{-3}$, respectively, assuming the typical radiation field $U=1$ and $\gamma=0.3$ for molecular clouds. The enhanced local radiation field near Cas~A can help smaller grains (i.e., results in a smaller $a_{\rm align}$) to be aligned by RATs, but higher density regions would require larger grains in order to be aligned.
Therefore, the grains required to produce the observed polarization in Cas~A must be large as grains are aligned via radiative torques, with grain size $a$ greater than $a_{\rm align}\sim 0.1\mu$m.

\cite{hoang16} showed that the alignment of grains with the magnetic field by RATs can be enhanced by paramagnetic relaxation (aka. Davis-Greenstein mechanism, \citealt{Davis_Greenstein51}). The characteristic timescale of paramagnetic relaxation is given by 
\begin{eqnarray}
\tau_{\rm mag} &= &\frac{I_{\|}}{VK(\omega)B^{2}}=\frac{2\rho~ a^{2}}{5K(\omega)B^{2}}\nonumber\\
&
\simeq& 0.34~\frac{\hat{\rho}~a^{2}_{-5}~T_{\text{d,1}}}{\hat{p}}\left(\frac{B}{10\,\rm m G}\right)^{-2}\left(1+(\omega\,\tau_{\rm el}/2)^{2}\right)^{2} ~{\rm yr,}\label{eq:tau_DG}~~~
\end{eqnarray}
where $I_{\|}$ is the principal inertia moment, $V$ is the grain volume, $K=\omega \chi"$ with $\chi"$ is the imaginary part of grain magnetic susceptibility, $T_\text{d,1}=T_\text{d}/10~{\rm K}$, $\hat{p}=p/5.5$ with $p$ as the coefficient of atomic magnetic moment, and $\tau_{\rm el}\sim 10^{-12}/f_{p}$ with $f_{p}$ the fraction of Fe atoms in the silicate material is the electron spin-spin relaxation time (see \citealt{hoang16} for details). 
       
From Equations (\ref{eq:tgas}) and (\ref{eq:tau_DG}) one can see that paramagnetic relaxation is much faster than gas randomization. Therefore, the combined effects of paramagnetic relaxation and RATs (aka. the M-RAT mechanism; \citealt{hoang16}) can help grains with size $a>a_{\rm align}$ to achieve the perfect alignment of the grain axis of maximum inertia moment with the ambient magnetic field (\citealt{hoang16}).
More precisely, because the grains spin along the shortest axis \citep[due to internal relaxation;][]{purcell79}, the spin (i.e., the grain angular momentum) is aligned with the magnetic field (\citealt{Hoang.2022}).

\subsubsection{Origin of the depolarization}
\label{Sdepol}

Our analysis of grain alignment in the previous section suggests that grain alignment in Cas~A is very efficient due to the combination of RATs and paramagnetic relaxation. The question now is what is the origin of the depolarization, as shown by $p-I$ anti-correlation (Fig.~\ref{polintrelation})?

In general, the observed decrease of polarization ($p$) with the intensity ($I$), can be induced by the decrease of the grain alignment efficiency and the tangling of the magnetic field along the line of sight (e.g., \citealt{seifried19,pattle2019,hoang21}). Below, we discuss in detail these two possibilities.

First, the polarization fraction of thermal dust emission depends on the size distribution of aligned grains with lower cutoff $a_{\rm align}$ and upper cutoff $a_{\rm max}$ \citep{Lee2020}. For a given $a_{\rm max}$ of the grain size distribution, the RAT alignment theory implies the value of $a_{\rm align}$ increases with the gas density $n_{\rm H}$ (see Eq. \ref{eq:aalign}), which narrows the size distribution of aligned grains and results in a decreasing polarization fraction. In dense regions eg $n_{\rm H}>10^{4}{\rm cm}^{-3}$ such that $a_{\rm align}> 0.3$ \mic, grain alignment is completely lost when $a_{\rm max}<a_{\rm align}$. As a result, thermal dust emission from this dense region would be unpolarized. Therefore, if $I \propto n_{\rm H}$, the RAT theory can reproduce the observed anti-correlation of $p-I$ with slope $\alpha \sim -1$.

Second, to check whether the magnetic field tangling can explain the depolarization, we analyze the dependence of the polarization fraction on the polarization angle dispersion function ($S$). The definition of $S$ was introduced in \cite{planck15}. The de-biased dispersion at position $\boldsymbol{r}$ on the sight-line is given as $S(\boldsymbol{r}) = \sqrt{S^{2}_{\rm biased}(\boldsymbol{r})-\sigma^{2}_{S}(\boldsymbol{r})}$ with $S_{\rm biased}(\boldsymbol{r})$ and $\sigma_{S}^{2}(\boldsymbol{r})$ the biased dispersion and its associated error at this position. They are respectively determined as
\begin{eqnarray}
    S_{\rm biased}(\boldsymbol{r}) &=& \sqrt{\frac{1}{N}\sum^{N}_{i=1}[\psi(\boldsymbol{r}+\boldsymbol{\delta}_{i})-\psi(\boldsymbol{r})]^{2}} \label{eq:S} \\
    \sigma_{S}^{2}(\boldsymbol{r}) &=& \frac{1}{N^{2}S_{\rm biased}^{2}}\sum_{i=1}^{N}\sigma^{2}_{\psi}(\boldsymbol{r}+\boldsymbol{\delta_{i}})[\psi(\boldsymbol{r}+\boldsymbol{\delta}_{i})-\psi(\boldsymbol{r})]^{2} \nonumber \\
    &&+\frac{\sigma^{2}_{\psi}(\boldsymbol{r})}{N^{2}S_{\rm biased}^{2}}\left[\sum_{i=1}^{N}\psi(\boldsymbol{r}+\boldsymbol{\delta}_{i}) - \psi(\boldsymbol{r})\right]^{2}
\end{eqnarray}
where $\psi(\boldsymbol{r})$ and $\sigma(\boldsymbol{r})$ are the polarization angle and its error at position $\boldsymbol{r}$, $\sigma(\boldsymbol{r}+\boldsymbol{\delta})$ are for $\boldsymbol{r}+\boldsymbol{\delta}$, $N$ is all data points within a circular aperture centered at the position $\boldsymbol{r}$ with a diameter of two beam sizes. 

Observations and numerical simulations of dust polarization from molecular clouds reveal the anti-correlation of the polarization fraction with the polarization angle dispersion $S$ \citep[e.g.,][]{planck15, planck15b}. This anti-correlation is expected because larger $S$ corresponds to stronger magnetic field tangling, which reduces the net polarization fraction. Here, using the observed polarization data, we compute the polarization angle dispersion using Equation~\ref{eq:S}.  
Figure~\ref{polparam} shows the variation of the polarization fraction with the polarization angle dispersion observed in Cas A. For small angle dispersions of $S$ $<$ 30 degrees, there is no clear correlation between $p$ and $S$  (see blue symbols in Figure~\ref{polparam}). However, for the large polarization angle dispersions of $S$ $>$ 30 degrees (red symbols in Figure~\ref{polparam}), the polarization degree decreases with the increasing $S$. This suggests that the main process for the depolarization is the loss of grain alignment, and the effect of B-field tangling becomes important for sufficiently large angles of $S>30$ degrees.


Therefore, the depolarization in $p-I$ anti-correlation is likely caused by the decrease of grain alignment toward denser regions, and the tangling of the magnetic field plays a minor role. The dense region observed in the SE shell (region C in Figure \ref{polpercentregions} (top)) indeed shows weaker polarization whereas the less dense region (labelled `diffuse', region B) shows a stronger polarization fraction (Table \ref{Tpol}).

\subsection{Dust properties in Cas~A}
We compare the observed polarization properties of Cas~A with a theoretical calculation that we described in Section 3, and we summarize what we have learnt about the dust properties in Cas~A below.

\noindent $\bullet$ The dust grains are large, indicated by the fact that a high polarization fraction is detected in far-IR emission \citep[see Section \ref{sec:alignment2};][]{guillet18}. Equation (\ref{eq:aalign}) implies that grains must be larger than $\sim$0.15 \mic\ for $n_{\rm H}$ $>$ 10$^3$ cm$^{-3}$ to reproduce the high polarization level. The suggestion of large grains in Cas~A is in agreement with observations of dust in the Crab Nebula \citep{gomez12b,owen15}, and these may be more robust to destructive processes in SNRs \citep{priestley22}. 

\noindent $\bullet$ The dust grains in Cas~A have elongated shapes. If there are only spherical grains, there will be no polarization \citep{kirchschlager19a, draine21}. The required elongation depends on the alignment efficiency.
If grain alignment is very efficient (e.g., perfect efficiency), a moderate elongation of the axial ratio of 1.4 is sufficient. The elongation of grains is likely greater than the axial ratio of 1.4. The condition of Cas~A is indeed unique because of the strong magnetic field and RATs, under which the grains can easily produce perfect alignment. However, the exact elongation of dust grains requires the detailed modeling of dust polarization, taking into account the grain alignment efficiency and magnetic field geometry.

\noindent $\bullet$ The high polarization levels observed in this work can be used to indicate the dominant grain composition responsible for the polarized emission. \cite{vandenbroucke21} simulated the
maximum linear polarization fraction as a function of wavelength for a dust grain mixture of elongated silicate grains and non-aligned graphite
grains assuming an MRN size distribution \citep{mathis97,zubko04}. They found that silicate grains can produce polarization levels from 5-15\,per\,cent \citep{guillet18,vandenbroucke21} whereas the polarization of carbon dust is less than 3\,per\,cent at 154 \mic\
\citep[Fig.~13 of][]{guillet18}. Pure carbon dust is not expected to be efficiently aligned and cannot produce the higher polarization levels observed in the HAWC+ images\footnote{Moreover, SOFIA/HAWC+ observations toward a C-rich AGB star, IRC+10216, reveal a far-IR polarization of less than 5\,per\,cent \citep{Andersson.2022}.} \citep{Chiar:2006,lazarian15,vandenbroucke21}.  A polarization fraction of (25, 20, 10)\,per\,cent requires (100, 80, 40)\,per\,cent silicate dust, respectively whereas a combination of 60\,per\,cent silicate and 40\,per\,cent graphite grains would
produce $\sim$ 15\,per\,cent polarization \citep{vandenbroucke21}. The levels of polarization vary across the SNR, areas with lower polarization levels (ie regions A and C) perhaps contain more carbon dust than those with more strongly polarized emission (ie region B, the diffuse area between the bright dust clouds). Alternatively, other SN grain compositions, such as iron-bearing dust may explain the origin of polarization levels higher than 25\,per\,cent. Metallic iron-type dust (pure Fe, FeS, etc.) are both observed \citep{rho08, arendt14} and predicted in SNe ejecta \citep{marassi19, sarangi15, sluder18}.  Unfortunately, there are no polarization simulations of Fe-bearing dust to which we can compare. Future dust polarization modeling, including Fe, FeS, and SiO$_2$ grains, and in the environment of shocks is encouraged in order to compare with the high polarization levels seen in Cas~A. 

\begin{figure*}
\includegraphics[width=15.2truecm]{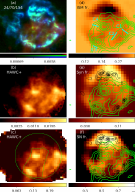} 
\caption{Three mosaicked colour images of \spitzer\ 24 \mic\ (in blue with the range from 100 - 1100 MJy sr$^{-1}$), Herschel 70 \mic\ (in green with the range from 0 -- 0.068 Jy pixel$^{-1}$ and pixel size is 1$'$), and HAWC+ 154 \mic\ (in red with the range from 0 - 0.021 Jy pixel$^{-1}$ and pixel size is 2.75$'$). (b) HAWC+ 154 \mic\ image in full resolution (13.6$''$). (c) HAWC+ 154 \mic\ image after rebinning by a factor of 3$\times$3 (to 41$''$ resolution). 
The HAWC+ 154 \mic\ contours (in green) are superposed on 
the fraction of ISM (d), synchrotron (e), and SN dust (f) maps from \citet{deLooze17}. 
The MIPS 24 \mic\  contours (in black) are on (e) and (f) maps. 
The fraction of ISM for regions around Cloud 1 (C1), C2, C3, and C4 is 0.58, 0.55, 0.65, and 0.93, respectively. The average fraction of ISM in the 160 \mic\ emission was estimated to be 0.65--0.7 of the total flux.
The images are centered on R.A.\ $23^{\rm h} 23^{\rm m} 27.75^{\rm s}$ and Dec.\ $+58^\circ$48$^{\prime}47.75^{\prime \prime}$ (J2000) with each FOV of 6.03$'$x6.03$'$.}
\label{sixpanelimages}
\end{figure*}

\begin{figure}
\includegraphics[scale=0.6,angle=0,width=8.1truecm]{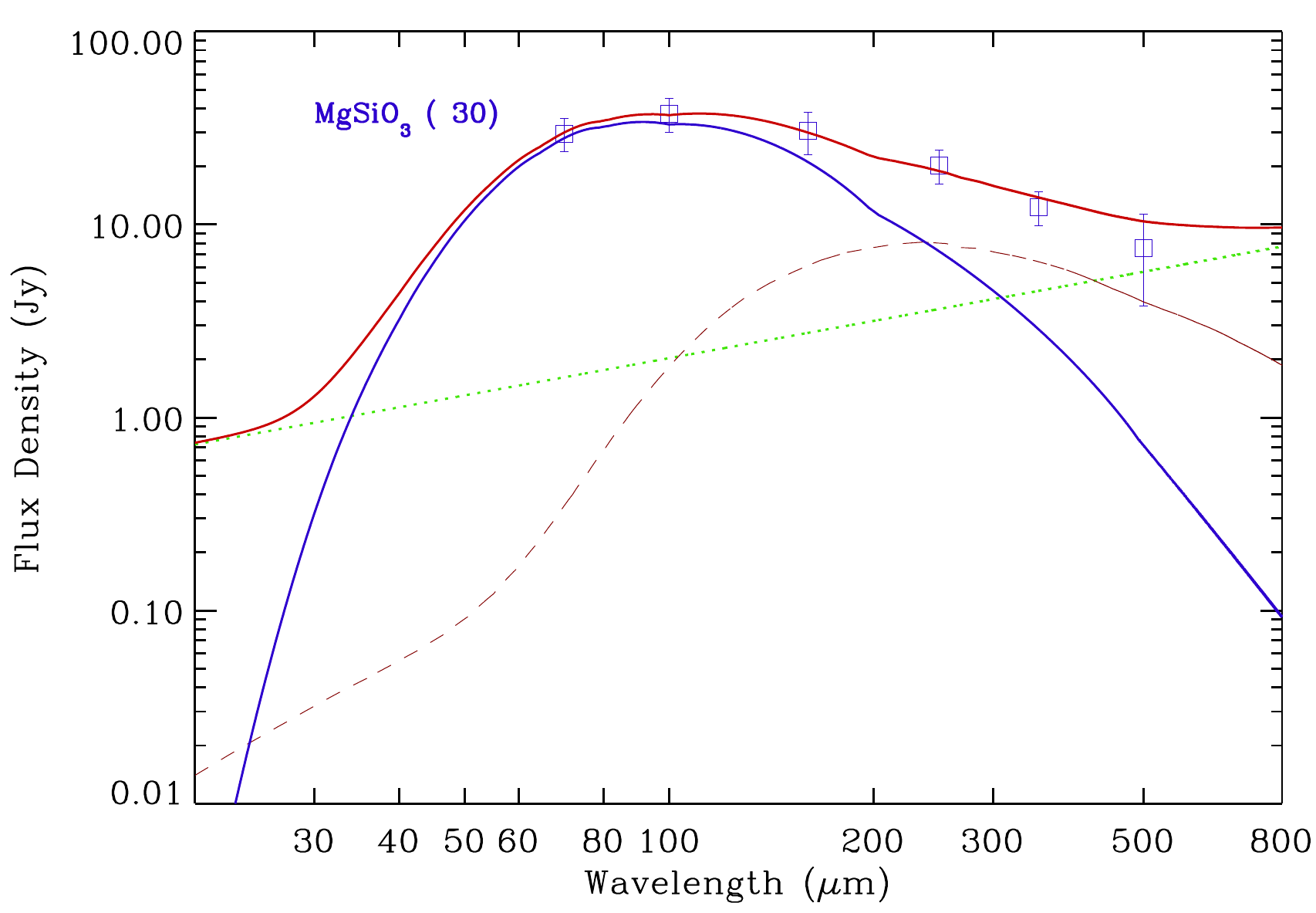} 
\caption{
The mid- and far-IR SED of the polarization region (the polygon in Fig. \ref{sixpanelimages}c). The \herschel\ photometry (squares) is superposed on the best-fit SED (blue) derived using the CDE model with grain composition MgSiO$_3$. The number in parenthesis is the dust temperature of the grain in K. The non-thermal synchrotron radiation is shown by the green dotted line, and thermal emission from IS dust is shown by the brown dotted line. The total flux density from the model fit is shown by the red, solid line.}
\label{poldustSED}
\end{figure}

The high polarization levels in this work therefore indicate the grains are silicate dominated. We caution however that polarization observations from a single band is insufficient to understand the polarization alignment mechanism and quantify the dust properties more accurately, for example, to estimate the exact contribution of Si/C and Fe dust. Future multi-band far-IR observations should break the degeneracies in grain composition and size, magnetic field conditions, and grain alignment efficiencies in the aftermath of the stellar explosions of massive stars.

\subsection{SN-dust mass and polarized far-IR emission}
\label{Sdustmass}

\begin{table*}
\begin{center}
\begin{tabular}{lllll}
\hline  \hline    
Instrument & Waveband (\mic)     & Total minus bkg (Jy)$^a$&  POL region$^c$ (Jy)  \\ \hline
\herschel\ PACS & 70 \mic\  &74.71$\pm$9.0 & 29.59$\pm$5.69 \\
\herschel\ PACS & 100 \mic\  &79.88$\pm$10.93 & 37.42$\pm$7.45  \\
HAWC+ & 154 \mic\  &47.52$\pm$ 7.89$^b$  & 26.12$\pm$6.11  \\
\herschel\ PACS & 160 \mic\ & 60.05$\pm$11.75 & 30.58$\pm$7.55  \\
\herschel\ SPIRE & 250 \mic\ & 45.49$\pm$0.41 & 20.21$\pm$0.25 \\
\herschel\ SPIRE & 350 \mic\ & 29.74$\pm$0.52 & 12.33$\pm$0.32 \\
\herschel\ SPIRE & 500 \mic\  & 19.35$\pm$0.26 & 7.55$\pm$0.15\\
\hline \hline
\end{tabular}
\end{center}
\caption{Far-infrared Flux Densities of HAWC+ and Herschel Images. $^a$The flux density of Cas~A with background subtraction. The chop-and-nod background region for 160 \mic\ background subtraction is used.
$^b$The total is for the region covered by HAWC+, marked in Fig.~\ref{sixpanelimages}(c) as a circular region. $^{c}$The POL region contains strong polarization vectors and the exact region is marked as a polygon in Fig. \ref{sixpanelimages}c.}
\label{Tphotometry}
\end{table*}

In this Section, we derive the mass of dust associated with the SNR Cas~A using the HAWC+ observations. We use the region containing the polarization vectors in Figure \ref{Bpolmap} (the exact region is marked as a polygon in Figure \ref{sixpanelimages}c; hereafter we call it the POL region.)

There are three main issues to consider when estimating the mass of dust: (i) removing the synchrotron contamination at IR wavelengths, (ii)  removing the contribution from interstellar material along the line of sight, and (iii) the assumed dust grain composition.

The synchrotron contribution is estimated using the spectral index $\alpha$=-0.64 \citep{rho03} where flux $S$ is proportional to frequency $\nu$ as log $S \propto$ $\alpha$ log $\nu$. The small spectral index variation \citep{rho03} across the Cas A SNR was ignored. This estimate of synchrotron emission is consistent with that of  \cite{dunne03}, where they found synchrotron made up more than half of the emission at 850 \mic\ and of \cite{deLooze17}, where they show synchrotron makes up 11$\pm$3.3 \,per\,cent of the 160 \mic\ emission. The synchrotron contribution at 154\mic\  is expected to be less than 8\,per\,cent (see Section \ref{S4.4.1} and Figure \ref{poldustSED} for details).

Removing the contribution from interstellar material along the line of sight has been an issue for Cas A. \cite{krause04} had originally suggested that no cold dust existed within Cas~A because part of the SNR  coincides with large interstellar CO clouds based on low-resolution CO data. However, the \spitzer\ MIPS 70 \mic\ map reveals dust emission following the morphology of the Cas~A SNR and bounded by the forward and reverse shocks \citep{rho08}. Most importantly, \herschel's high-resolution images revealed far-IR emission from dust associated with the ejecta \citep{barlow10, deLooze17}.
\cite{deLooze17}  derived a suite of insterstellar models for Cas A by assuming an SED shape appropriate for interstellar emission at IR wavelengths for different radiation field strengths, $G_0$, and dust composition. There is an inherent uncertainty in this method due to the assumption of the SED shape and uncertainty in $G_0$ but these models represent the best spatial separation of the contaminants to the SN-dust emission signal. Model maps from \citet{deLooze17} are shown in Figure \ref{sixpanelimages}d - \ref{sixpanelimages}f alongside the HAWC+ observations. The fraction of the IR flux across Cas~A originating from unrelated foreground ISM emission is shown in Figure \ref{sixpanelimages}d, similarly the fraction of the emission from synchrotron emission is shown in Figure \ref{sixpanelimages}e. Combining these maps, \citet{deLooze17} created a map of the fraction of the emission originating from SN dust (Figure \ref{sixpanelimages}f). These indicate that the average fraction of SN, IS and synchrotron emission over the total SNR at 160 \mic\ is 0.28, 0.7 and 0.05-0.2 respectively \citep{deLooze17}.

Three-colour images of \spitzer\ at 24 \mic, \herschel\ at 70 \mic, and
HAWC+ at 154 \mic\ show where the 154 \mic\ excess emission resides (in red
in Fig.~\ref{sixpanelimages}a).  The excess emission (e.g., outside the SNR
on the west) is consistent with the interstellar fraction map in Figure
\ref{sixpanelimages}d, except for the northern shell where far-IR emission
at 154 \mic\ is lacking. (Note that the full resolution HAWC+ image
(13.6$''$) at 154 \mic\ is comparable to that (12$''$) of \herschel\ 160
\mic.) Four interstellar clouds are identified in the ISM fractional map
from \citet{deLooze17} and are marked in Figure \ref{sixpanelimages}d. The
bright interstellar cloud observed to the east of the center in Figure
\ref{sixpanelimages}d (marked with a line as cloud 4), is offset from the
bright knot of cold dust emission at the center. The bright IS cloud in the
south (cloud 3) is also not coincident with the southeastern shell observed
at 154 \mic. This suggests that there are dusty regions in the 154 \mic\
observation of Cas~A that are not dominated by foreground IS dust. The
western side of the remnant has ISM dust where there is 154 \mic\ emission
but no polarization is detected (red emission is noticeable outside the SNR
shell in Figure \ref{sixpanelimages}a).

Now with the polarization map at 154 \mic, what can we say about the dust mass associated with SN-dust and the contribution from ISM-dust? The average polarization of the dust is 19.4$\pm$6.7\,per\,cent. The western shell of Cas~A shows a lack of polarization, which is consistent with this region being contaminated by the ISM  \citep{krause04}. The strong polarization in far-IR detected in this work (see Figs \ref{Bpolmap} and \ref{polpercentregions}) indicates the presence of SN-dust in Cas~A and not just ISM dust towards the SNR. We detect polarization at the center of the SNR, in the southeastern shell, and in between these two regions (Regions A, B, and C in Fig.~\ref{polpercentregions}). Comparing with the \citet{deLooze17} model maps, the polarized areas are coincident with regions with higher fractions of SN-dust (about 50\,per\,cent).

We remeasured the \herschel\ fluxes to confirm that our measurements are consistent with those in Table 2 of \cite{deLooze17}.
We estimated the total flux density of Cas A within a large area (20 \% larger than the bright outer shell of Cas A) at 160 \mic, which is consistent with the value 236.2$\pm$19.2 Jy obtained by \cite{deLooze17}. Now though with background subtraction (using the chop-nod area shown in Fig. \ref{hawcpfootprint}), the flux of the total Cas A SNR (we define a region that encompasses the area covered by the HAWC+ observation which is marked as a large circle shown in Fig.~\ref{sixpanelimages}c)
reduces to 60.1$\pm$11.8 Jy (Table~\ref{Tphotometry}). 
In comparison, the HAWC+ flux at 154 \mic\ is 47.5$\pm$7.9 Jy. 
These fluxes are within the range of SN dust estimated using the model of \citet{deLooze17} (132$\pm$16, 70$\pm$12, and 25$\pm$7 Jy for $G$ = 0.3 $G_0$, 0.6 $G_0$, and 1.0 $G_0$ respectively.

\subsubsection{The Dust Mass within the Polarized Region}
\label{S4.4.1}

To estimate the amount of SN dust in Cas~A we performed photometry of 
the POL region (the polygon in Figure \ref{sixpanelimages}c).
The shape was chosen to exclude the western region and also avoid the ISM dominated areas, guided by the ISM fraction map from \cite{deLooze17} and the three-colour maps in Fig. \ref{sixpanelimages}a. 
The POL flux density is 26.12$\pm$5.11 Jy at 154 \mic\ with HAWC+ and 30.58$\pm$7.55 Jy at 160 \mic\ with \herschel. The flux density of the POL region is 55$\pm$14\,per\,cent of the emission from the entire SNR covered by the HAWC+ observation (the large circle in Fig.~\ref{sixpanelimages}c). The spectral energy distribution of the POL region using \herschel\ and SOFIA/HAWC+ is shown in Fig. \ref{poldustSED} with fluxes listed in Table \ref{Tphotometry}.

Since the polarization indicates silicate dust grains dominate, we use the MgSiO$_3$ non-spherical CDE model (since this provides an elongated grain shape with the axis ratio greater than 1.4).\footnote{We note that silicate grains of MgSiO$_{3}$ structure are paramagnetic material, which can align with the magnetic field.} The Q$_{\rm abs}$/$a$ of the CDE model is only 2-3\,per\,cent higher than that of the spherical model using silicate dust of MgSiO$_3$ and it retains the same shape above 10 \mic. Here we have assumed a grain size $a = 0.15$ \mic. Note that we excluded SiO$_2$ or Mg$_{0.7}$SiO$_{2.7}$ grains that were previously shown to produce the sharp 21 \mic\ dust feature in Cas~A \citep{rho08, rho18} 
because the IRS spectrum for the POL region we extracted did not show this feature.
We first performed dust spectral fitting, resulting in a best fit dust temperature below 20 K, which indicates there is still a potential contribution from IS cold clouds \citep{reach95, boulanger96, lagache98, millard21}. We then included the ISM SED model from \cite{jones13} and fit the SED simultaneously (Fig. \ref{poldustSED}) using SN dust composed of MgSiO$_3$ and ISM SED assuming G$_0$=0.6 (as was done by \cite{deLooze17}). We estimated the synchrotron emission using the 4.72 GHz flux density \citep{anderson91} and radio spectral index of 0.64 \citep{deLooze17} based on the Planck emission from Cas~A \citep{deLooze17, planck31arnaud16}. The fitting yielded a SN dust temperature of MgSiO$_3$ of 30$\pm$3 K, and a SN dust mass of 0.13$\pm$0.1 M$_\odot$.
We repeated the same steps for $G=0.3~G_0$ and $G=1 G_0$ and estimated a SN dust mass of 0.15$\pm$0.3 M$\odot$ and 0.12$\pm$0.02 M$\odot$, respectively. These masses (0.1 - 0.18 M$_\odot$) serve as the lower limit of the dust mass since we believe a high-resolution polarization map will reveal polarization on smaller scales outside the POL region.

The POL region dust mass (0.14$\pm$0.04 M$_\odot$, hereafter) is roughly a 1/3 of the total dust mass observed previously in Cas~A \citep{dunne09,deLooze17}. It is however consistent with the lowest dust mass estimate proposed by \citet{deLooze19} when assuming a high ISRF ($G = 1.0 G_0$). Our estimate is less dependent on the radiation field strength since the ISM contribution is lower in the POL region compared to other parts of the SNR. 

Are such high dust masses expected in SNe ejecta? Theoretical models of SN dust formation show that the mass formed depends on the progenitor mass and the SN type \citep{nozawa07,todini01,hirashita14}. The dust mass predicted to form from a 20-25 M$_\odot$ progenitor star is about 0.5 M$_\odot$ at solar abundance \citep{todini01}. The dust mass produced by SN does depend on the density of the environment, for example, only 0.1 -- 0.2 M$_\odot$ of dust is predicted to form when the density is $\rm n_H$ = 1 cm$^{-3}$ \citep{hirashita14}.

\subsubsection{Implications for dust in the early Universe}

Models that follow the dust production rates of galaxies in the early universe (500-800 Myrs after the Big Bang) require 0.08 -- 0.3 M$_\odot$ of dust \emph{per core collapse SN} at zero metallicity, increasing to 0.24 -- 0.9 M$_\odot$ at solar metallicity to explain the large quantities of dust observed \citep{morgan03,dwek04,gall11,michalowski15}.  The amount of dust that is required to form is both dependent on the initial mass function assumed in the models and the fraction of the dust that survives.  In models assuming a top-heavy IMF, less dust is required per ccSNe than for a Salpeter IMF due to the higher fraction of high mass stars. For dust survival, since the polarization levels observed with HAWC+ implies large grains within the SNR ($>$ 0.15 \mic), grain destruction processes are expected to be less efficient, and thus a significant mass of SN dust may survive the harsh environment of the SNR. The $\sim$0.14 M$_\odot$ of SN dust estimated from the polarized region is still a significant amount to account for the dust in the early Universe, eg this could explain the origin of dust in 4 out of the 6 high-$z$ galaxies modelled by \citet{michalowski15} (see their Fig.~1).

\section*{Conclusions}
\begin{enumerate}
\item The polarization map obtained with SOFIA/HAWC+ at 154 \mic\ differs from the previous polarization map at 850 \mic\ \citep{dunne09}.
The high polarization observed in the northern shell in the 850 \mic\ map is faint at 154 \mic. This may be due to an over-subtraction in this region due to the chop-nod process. Elsewhere, the 154 \mic\ map largely follows the distribution of cold dust emission, with bright emission at the southeastern shell and the center.

\item We measure 5-30\,per\,cent polarization across Cas~A indicating that SNRs may be a strong polarization source in
the far-IR. The high polarization is direct evidence that grains are
elongated and aligned to a high degree. 
The detection of highly polarized dust emission in Cas~A supports the hypothesis that significant quantities of SN dust have formed and suggests that a large fraction of the 154 \mic\ emission is from SN dust grains.

\item We observe an inverse correlation
between the intensity and polarization degree. The polarization shows that grains are aligned but at a weaker polarization level where the emission is bright at 154 \mic, including the central unshocked ejecta and the southeastern shell. In contrast, stronger polarization is detected in regions between the bright structures. 

\item Using the modern theory of grain alignment, we find that silicate grains can be perfectly aligned due to the joint action of paramagnetic relaxation and radiative torques due to the large magnetic field strength in Cas~A. This can explain the high polarization fraction observed by SOFIA/HAWC+. Moreover, the efficiency of grain alignment by  RATs decreases with gas density.

\item The depolarization in the $p-I$ anti-correlation with a steep slope of $\alpha=-1$ can be reproduced by the RAT theory - grain alignment is efficient in the outer regions due to its low gas density. The grains required to produce such anti-correlation of $p-I$ must be large as grains are aligned via radiative torques, with grain size $a$ greater than $a_{\rm align}\sim 0.1 \mu$m.
However, in very dense regions, grain alignment is lost where the minimum grain size required for RAT alignment to occur exceeds the maximum grain size present, meaning that the grain sizes in the very dense region may not be large.

\item We performed photometry of the polarized region (the polygon region in Fig. \ref{sixpanelimages}c) and spectral fitting of its spectral energy distribution (SED) using the dust properties implied by the polarization. We assume silicate dust, Enstatite (MgSiO$_3$) using the CDE model accounting for elongated and large grains. The dust spectral fitting yielded a dust temperature for MgSiO$_3$ grains of 31 K, with $\sim$0.14 M$_\odot$ of mass. 
This mass serves as the lower limit of the dust mass produced in the Cas~A ejecta; this is still a significant amount to account for the dust in the early Universe. 

\end{enumerate}

Future JWST observations, with its higher spatial and spectral resolution,
will help to accurately constrain gas density, velocity dispersion, and the
warm dust temperature, which will lead to a more accurate estimate of
magnetic field strengths in SNRs. High-resolution far-IR observatories with
polarization capabilities are key to accurately estimating dust masses and
determining the role of SN-dust in the early Universe. Unfortunately, SOFIA
had its last flight on 2022 September 28 and the SPICA and Cosmic Origins
programs have been canceled. There is currently no far-IR observatory or
space mission on the horizon that will  challenge us to answer this
fundamental astrophysical question of the origin of dust in the early
Universe. Far-IR polarization measurements are vital to answer the question
since some of the key uncertainties in dust properties needed to constrain
the dust mass can be resolved.

\section*{Acknowledgements}
This research is based on observations made with the NASA/DLR Stratospheric Observatory for Infrared Astronomy (SOFIA). SOFIA is jointly operated by the Universities Space Research Association, Inc. (USRA), under NASA contract NNA17BF53C, and the Deutsches SOFIA Institut (DSI) under DLR contract 50 OK 2002 to the University of Stuttgart. This research has also made use of the NASA/IPAC Infrared Science Archive (IRSA), which is funded by the National Aeronautics and Space Administration and operated by the California Institute of Technology.
We thank the anonymous referee for insightful comments.
APR would like to thank Sangwook Park for insightful discussions on improving signal to noise ratio in our polarization measurements. MM thanks the HAWC+ instrument team and SOFIA staff for their hospitality during the SOFIA flight.
JR and APR acknowledge support from NASA through the award SOF07\_0047 issued by USRA and  the ADAP award 80NSSC20K0449.
TH is supported by the National Research Foundation of Korea (NRF) grant funded by the Korea government (MSIT No. 2019R1A2C1087045). 
MB and FK acknowledge support from ERC Advanced Grant 694520 SNDUST.

\section*{Data Availability}

 The re-binned data products are to be available through VizieR Online Data Catalog and upon request to the first author. The SOFIA/HAWC+ data products presented here are also available through the the Caltech/IPAC IRSA website.
\bibliographystyle{mnras}
\bibliography{msrefsall}

\appendix
\section{Chop-Nod Simulation}
\label{AppSimu}
We briefly investigate the bias introduced by the Chop-Nod strategy implemented in the SOFIA/HAWC+ pipeline.
We note that the Chop-Nod method was the only available mode with SOFIA/HAWC+ at the time of this observation. The following investigation is meant to warn the reader about potential uncertainties.

The main issue related to the Chop-Nod strategy is the removal of a \textit{flat} ISM in a region that can significantly fluctuate. This may lead to oversubtraction in pixels that do not have a lot of ISM contribution, and lead to losing that information.

This could be a part of the reason
why a portion of the region studied by Dunne et al. (2009), found to
have a high polarization fraction at the SCUBA wavelength, is now
showing negative Stokes I values (the upper part of the northern shell; the white contour and outside the contour in Fig.~\ref{FigNorth}) at 154 \mic. Another reason of lack of polarization at the northern shell may be because the intensity of the emission and polarization changes depending on wavelengths \citep{chastenet22, tram21}. The morphology of cold dust of far-IR emission intrinsically differs from those of the radio, 850 \mic\ and warm dust (24 \mic) emission as shown in Figs.~\ref{sixpanelimages} and \ref{FigNorth} and the degrees of the polarization are also different.

\begin{figure}
    \centering
    \includegraphics[width=0.5\textwidth]{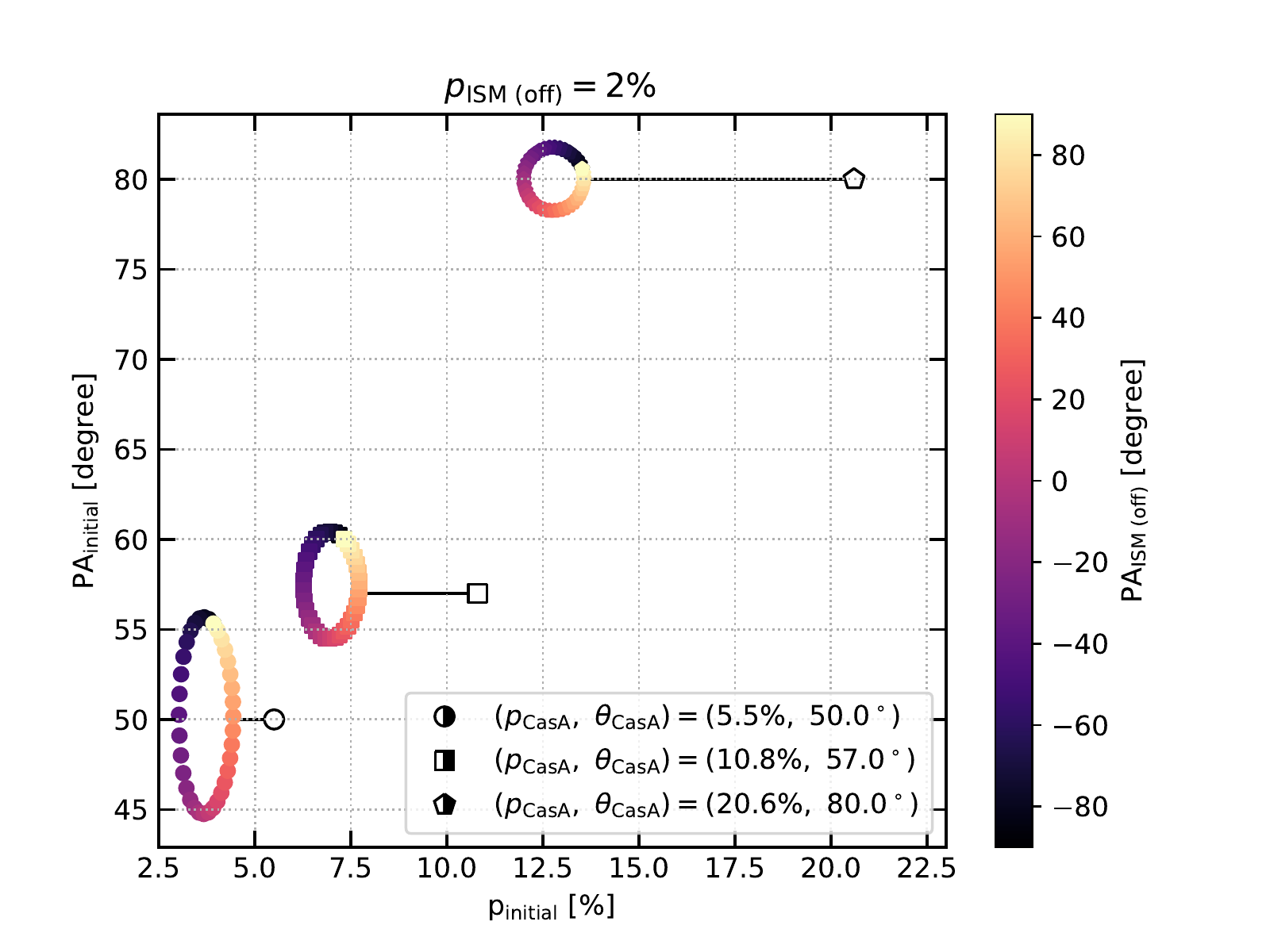}
    \caption{\textbf{Measured} target polarization (\textbf{empty} symbols)\textbf{, and the initial polarization before chop-nod subtraction}. The oversubtraction of a constant ISM value can artificially boost the resulting polarization fraction.}
    \label{FigSim}
\end{figure}

\begin{figure}
    \centering
    \includegraphics[width=0.4\textwidth]{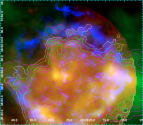} 
    \caption{The mosaicked image of Cas A focuses on the northern region; RGB represents HAWC+ 154 \mic, \herschel\ 160 \mic\ (with purple contours of 0.0028, 0.0033 and 0.0038 Jy/pixel where the pixel size is 1$''$), and radio (1.38 GHz) images, respectively. The contours of the HAWC+ image are marked in green (positive values of 0.0022 and 0.006 Jy/pixel where the pixel size is 2.75$''$ ) and white (negative value of -0.0001 Jy/pixel). The lower part of the northern shell agrees between HAWC+ and \herschel\ images. The upper part of the north shell shows a negative value (white contour) in the HAWC+ image, and the contours show a discrepancy between HAWC+ and \herschel\ images. Moreover, the HAWC+ and \herschel\ far-IR images show stronger emission at the inner part of the northern shell and are intrinsically different from the radio (emission in blue), 850 \mic\  \citep{dunne03, dunne09}, and warm dust (\spitzer\ 24 \mic) images on the upper part of the northern shell (see Fig.~\ref{sixpanelimages}a).}
    \label{FigNorth}
\end{figure}

Here we examine the effect of the polarization that exists in the surrounding ISM emission on that of the SNR Cas A when we use the chop-nod strategy. In Figure~\ref{FigSim}, we show the overestimation of a few intrinsic polarization values due to the subtraction of the ISM. 
The underlying strategy is to ``remove'' the Chop-Nod pipeline. 
We use the Level~4 data, and re-inject the flat ISM, i.e., $(I, Q, U)_{\rm initial} = (I, Q, U)_{\rm Cas A} + (I, Q, U)_{\rm ISM}$.
If this ISM contribution is overestimated, the final measured polarization, considered the one of the target, is biased toward higher values.
We proceed as follows:
\begin{itemize}
    \item We use the comparison of the Herschel/PACS 160 microns and the SOFIA/HAWC+ 154 microns to derive an average ISM intensity that was subtracted in the HAWC+ pipeline. We find that value to be roughly $I_{\rm ISM} = 0.0018$~Jy/as$^2$.
    \item For a value of the ISM polarization $p_{\rm ISM} = 2$ per cent and a range of ISM polarization angles $\theta_{\rm ISM} \in [-90, 90]$ degrees, we find the Stokes parameters $Q_{\rm ISM}$ and $U_{\rm ISM}$ by solving the system of equations:
    \begin{equation}
        \begin{split}
        p &= \sqrt{Q^2 + U^2}/I, \\
        \theta_{\rm p} &= 0.5 \times {\rm arctan}(U/Q)
\end{split}
    \end{equation}
    The value $p_{\rm ISM} = 2$ per cent was found by \citet{planck15}.
    \item We next sample the rebinned Cas~A data that pass the quality tests outlined in the main text. Here we search for pixels with 5, 10, and 20 per~cent polarization fraction, and extract their Stokes parameters.
    These are given as the legend in Figure~\ref{FigSim}. We \textbf{add} the $(I, Q, U)_{\rm ISM}$ to these vectors, and recalculate their \textbf{initial} polarization properties. 
\end{itemize}

In this Figure, the difference in the polarization properties before and after Chop-Nod subtraction is shown by the horizontal lines. The ratio $I_{\rm initial}/I_{\rm ISM}$ is the main driver of this difference. A more detailed study will be needed to fully quantify the extent of biases in the polarization properties due to subtraction of the ISM. 
Alternatively, scan mapping, such as the on-the-flying mapping method, can be used to remove the bias from the chop-nod method.

\label{lastpage}
\end{document}